# Field and Globular Cluster LMXBs in NGC 4278

G. Fabbiano

*Harvard-Smithsonian Center for Astrophysics, 60 Garden St., Cambridge, MA 02138*

gfabbiano@cfa.harvard.edu

N. J. Brassington

*Harvard-Smithsonian Center for Astrophysics, 60 Garden St., Cambridge, MA 02138*

L. Lentati

*School of Physics and Astronomy, University of Southampton, Highfield, Southampton, SO17 1BJ, UK*

L. Angelini

*Laboratory for X-ray Astrophysics, NASA Goddard Space Flight Center, Greenbelt, MD 20771*

R. L. Davies

*Sub-Department of Astrophysics, University of Oxford, Oxford OX1 3RH, UK*

J. Gallagher

*Department of Astronomy, University of Wisconsin, Madison, WI 53706-1582*

V. Kalogera

*Northwestern University, Department of Physics and Astronomy, Evanston, IL 60208*

D.-W. Kim

*Harvard-Smithsonian Center for Astrophysics, 60 Garden St., Cambridge, MA 02138*

A. R. King

*Theoretical Astrophysics Group, University of Leicester, Leicester LE1 7RH, UK*

A. Kundu

*Department of Physics and Astronomy, Michigan State University, East Lansing, MI 48824-2320*

S. Pellegrini

*Dipartimento di Astronomia, Universita' di Bologna, Via Ranzani 1, 40127 Bologna, Italy*

A. J. Richings

*School of Physics and Astronomy, University of Southampton, Highfield, Southampton, SO17 1BJ, UK*

G. Trinchieri

*INAF-Osservatorio Astronomico di Brera, Via Brera 28, 20212 Milano, Italy*

A. Zezas

*Harvard-Smithsonian Center for Astrophysics, 60 Garden St., Cambridge, MA 02138*

*University of Crete, Physics Department, PO Box 2208, 71003, Heraklion, Crete, Greece*

*IESL, Foundation for Research and Technology, 71110, Heraklion, Crete, Greece*

and

S. Zepf

*Department of Physics and Astronomy, Michigan State University, East Lansing, MI 48824-2320*

## ABSTRACT

We report a detailed spectral analysis of the population of low-mass X-ray binaries (LMXBs) detected in the elliptical galaxy NGC 4278 with *Chandra*. Seven luminous sources were studied individually, four in globular clusters (GCs), and three in the stellar field. The range of (0.3-8 keV) $L_X$ for these sources is $\sim 3-8\times10^{38}$erg s$^{-1}$, suggesting that they may be black hole binaries (BHBs). Fitting the data with either single thermal accretion disk or power-law models results in best-fit temperatures ~0.7-1.7 keV and $\Gamma\sim1.2-2.0$, consistent with those measured in Galactic BHBs. Comparison of our results with simulations allows us to discriminate between disk and power-law dominated emission, pointing to spectral/luminosity variability, reminiscent of Galactic BHBs. The BH masses derived from a comparison of our spectral results with the $L_X\sim T_{in}^4$ relation of Galactic BHBs are in the 5-15 $M_\odot$ range, as observed in the Milky Way.

The analysis of joint spectra of sources selected in three luminosity ranges ($L_X\geq 1.5\times10^{38}$ erg s$^{-1}$; $6\times10^{37}\leq L_X<1.5\times10^{38}$ erg s$^{-1}$; and $L_X<6\times10^{37}$ erg s$^{-1}$), suggests that while the high luminosity sources have prominent thermal disk emission components, power-law components are likely to be important in the mid and low-luminosity spectra. Comparing low-luminosity average spectra, we find a relatively larger $N_H$ in the GC spectrum; we speculate that this may point to either a metallicity effect, or to intrinsic physical differences between field and GC accreting binaries.

Analysis of average sample properties uncover a previously unreported $L_X - R_G$ correlation (where $R_G$ is the galactocentric radius) in the GC-LMXB sample, implying richer LMXB populations in more central GCs. No such trend is seen in the field LMXB sample. We can exclude that the GC $L_X - R_G$ correlation is the by-product of a luminosity effect, and suggest that it may be related to the presence of more compact GCs at smaller galactocentric radii, fostering more efficient binary formation.

## 1. Introduction

The study of LMXBs and their evolution is one of the original and still important fields of Galactic X-ray astronomy. This study is now being pursued in external galaxies thanks to the sensitive sub-arcsecond resolution and photometric capabilities of the *Chandra X-ray Observatory* (Weisskopf et al 2000). With *Chandra*, samples of LMXBs have been detected in several early-type galaxies, both associated with GCs and in the stellar field, providing unique insight in the spatial distribution, luminosity function, and GC association of LMXBs. These results have rekindled the debate on the relative importance of dynamical interactions in GCs versus evolution of native binaries in the galaxy's stellar field for LMXB formation (see review, Fabbiano 2006 and references therein).

Although it has been established that point-like sources in early-type galaxies have average spectral properties consistent with those of Galactic LMXBs (see Fabbiano 2006), detailed comparisons of the spectral/luminosity behavior of these sources with the characteristic spectral states of Galactic LMXBs (e.g. McClintock & Remillard 2006; Remillard & McClintock 2006), require deeper and repeated observations. Most of the *Chandra* observations of early-type galaxies have provided single epoch snapshots, and limited statistics. Our deep monitoring observations of the elliptical galaxies NGC 3379 and NGC 4278 (PI: Fabbiano) provide the needed data for these investigations (see catalogs of Brassington et al 2008, 2009; hereafter B08, B09). Brassington et al (2010, hereafter B10), report the analysis of variable luminous sources discovered in the GC-poor galaxy NGC 3379, and establish a new simulation-based method for the interpretations of these spectral results. The works of B10 suggests that that these luminous sources in NGC 3379 follow spectral variability patterns similar to those of Galactic LMXBs.

The GC-rich elliptical NGC 4278 (B09) provides a far richer population of X-ray sources than NGC 3379. The six separate *Chandra* ACIS-S pointings of NGC 4278, covering a time span of two years, led to the detection of 236 sources, down to a limiting sensitivity of $\sim 3.5 \times 10^{36}$erg s$^{-1}$, with widespread flux and spectral variability. Of these sources, 180 are found within the $D_{25}$ ellipse of NGC 4278; comparison with existing *HST* WFPC2 data led to the identification of 39 GCs sources, and 71 field LMXBs (sources with no optical counterpart). This galaxy, therefore, provides for the first time large samples of both GC and field extragalactic LMXBs, all at the same distance, with extensive time coverage. At a distance of 16.1 Mpc (Tonry et al 2001), these sources cover a large range of X-ray luminosities, from the $10^{36} - 10^{37}$erg s$^{-1}$ range luminosities typical of most of the Galactic LMXBs, to luminosities of $\sim 5 \times 10^{38}$erg s$^{-1}$, which exceeds the Eddington luminosity of a neutron star. (Note that Jensen et al 2003 report a distance of 14.9 Mpc; although we adopt the Tonry et al distance for consistency with B09, this choice does not affect in any way our conclusions.)

The NGC 4278 data allow a detailed comparison of flux and spectral variability of individual luminous GC and field sources, complementing the study of NGC 3379 (B10) and the comparison with Galactic LMXBs; given the distance of NGC 4278 and the number of source counts needed for a meaningful analysis, these luminous sources are all black hole (BH) binary candidates with $L_X \geq 1.5 \times 10^{38}$ erg s$^{-1}$. In addition, the richness of the NGC 4278 LMXBs populations, both in the stellar field and associated with GCs, allows for the first time a systematic exploration of spectral properties of the two samples, which may provide further insight on their nature and evolution.

The rich GC LMXB population of NGC 4278 allows the exploration of the properties of extragalactic GC sources in a galaxy where these sources are observed in a luminosity range covering that of the Galactic LMXBs. Some of these properties have been established in several studies, but typically addressing more luminous sources ($\geq 5 \times 10^{37}$erg s$^{-1}$). Widely accepted results based on the statistics of GC associations show that the occurrence of LMXBs is more likely in more luminous, and in red - high metallicity clusters (see Fabbiano 2006; Kundu et al 2007). Other properties are still under discussion, including a debated metallicity-dependent hardening of the spectral parameters of the blue GC LMXBs (Maccarone et al 2003; Kim E. et al 2006; Sivakoff et al 2008), which has been suggested as evidence for winds speeding the evolution of LMXBs in low-metallicity blue clusters (Maccarone et al 2004). With the NGC 4278 LMXB samples we will revisit these questions and open new ones. In particular, we explore possible dependencies of GC properties on the galactocentric radius ($R_G$), which have been suggested in the Milky Way (van den Bergh et al 1991; Bregman et al 2006), but have not been reported so far in extragalactic LMXB populations (e.g., Sivakoff et al 2007).

In this paper, we first describe the observations and the samples of sources used in the analysis (§2), we report the results of the detailed time-resolved spectral analysis of seven luminous LMXBs (four in GCs and three in the field §3), we examine the average spectral properties as a function of source luminosity and optical counterpart (§4), and we then discuss a correlation analysis of the GC sample (§5). The implications of our results are discussed in §6 and summarized in §7.

## 2. Observations and Analysis Samples

The data were all acquired with the *Chandra* ACIS-S3 CCD chip and reduced as described in B09, resulting in a co-added exposure of 458 ks. Table 1 gives the log of the observations, where Obs. No. is the number we will use to refer to a given observation in this paper, OBSID is the unique *Chandra* identifier, Date is the date in which the data were acquired, and $T_{eff}$ is the effective exposure time of the cleaned data (see B09).

The samples of GC and field LMXBs studied in this paper are extracted from the catalog of B09. To have reliable optical identification, we only considered *Chandra* X-ray sources detected within the $D_{25}$ ellipse of NGC 4278 in the field observed with the *HST* WFPC2 (Kundu & Whitmore 2001; B09). In this paper we will refer to individual B09 sources with 'S' followed by the B09 source number.

(1) Individual luminous sources - From the joint *Chandra-HST* data set, we extracted four GC and three field LMXBs with $L_X \geq 1.5 \times 10^{38}$ erg s$^{-1}$ (B09) for individual spectral analysis. Empirically, we find that a minimum of $\sim 400$ co-added source counts are needed for our spectral analysis. Given the distribution of source counts in B09, all the sources in this luminosity range satisfy this condition. This luminosity range includes sources at or exceeding the Eddington luminosity of a neutron star. Given the distance of NGC 4278, fainter sources do not yield enough counts for a meaningful analysis. To overcome this problem, we also extracted and analyzed cumulative spectral data for different subsamples of sources, as discussed below.

For completeness, we report in Appendix A the results of the spectral analysis of additional six sources, which satisfy the selection criterion for individual spectral analysis, but do not have optical counterpart, falling outside of the *HST* field of view.

(2) Samples for joined spectral analysis - We extracted the most extensive possible samples of field and GC LMXBs for joined spectral analysis. The field sample includes 59 of the 71 field LMXBs listed in B09 (sources within the *HST* WFPC2 field of view with no optical counterpart, or determined to have a low probability of being matched to a GC - 'excluded matches', see Section 2.6 of B09). The 12 excluded sources were either found within 10" of the nucleus, where the field is too crowded to allow a 'clean' extraction of spectral counts, or had count extraction regions partially overlapping with GC-LMXB regions (Note that in this and similar cases of overlapping sources, we carefully examined our data and reduced the source extraction regions if possible - see also § 3; however, in most cases the contamination would have been too large and these sources were excluded). No sources with a separation $< 1''$ from an optical object are included in the field sample. We also extracted a more conservative field sample, excluding all sources with a separation <2" from a possible optical counterpart. However, comparing the spectral properties of the field sample to this more conservative sample we find no significant differences in the results (although the results of the more conservative sample are of course not as tightly constrained due to the poorer statistics of the smaller sample.) We therefore use the field sample containing 59 sources in all further analysis.

The GC sample includes only 28 of the 39 GC-LMXBs identified in B09; the remaining 11 sources were excluded because their extraction regions overlapped with those of field LMXBs.

We divided each sample into three subsamples containing sources in three ranges of X-ray luminosities: $L_X \geq 1.5 \times 10^{38}$ erg s$^{-1}$, which includes the sources also analyzed individually; $L_X$ in the range $6 \times 10^{37} - 1.5 \times 10^{38}$ erg s$^{-1}$; and $L_X < 6 \times 10^{37}$ erg s$^{-1}$. The latter two groups were chosen to roughly equalise the number of counts for spectral analysis. This procedure resulted in 4, 7 and 17 sources in the high, mid, and low-luminosity subsamples respectively for the GC LMXBs. Because of count extraction region overlaps among sources in different luminosity subsamples, of the 59 field sources only 51 were included in the subsamples resulting in 3, 7 and 41 sources in the high, middle, and low luminosity groups, respectively (e.g. S139 and S137 belong to the mid and low luminosity groups respectively, and overlap with one another, and so to maintain a clean source sample both were excluded at this stage, but these sources are both in the 'full' field sample, together with another six sources in a similar situation).

(3) GC color - To explore a reported, but controversial spectral dependance on metallicity (Maccarone et al 2003; Kim E. et al 2006), the GC-LMXBs were also subdivided based on the color of the host GC. For this purpose we used the V and I band data gathered from the *HST* observation of NGC 4278 to classify GCs with $V-I > 1.05$ as red, and those with $V-I \leq 1.05$ as blue (see Kundu et al 2007, for examples of bimodal $V-I$ GC distributions; also review Brodie & Strader 2006). After exclusion of sources partially overlapping field-LMXBs, the red group contained 17 sources and the blue group 11. The high luminosity sources are divided evenly between these two groups.

(4) Correlation analysis - For statistical work not involving spectral analysis, we considered a larger GC-LMXB sample, including all 39 of the GC-LMXBs in B09 for trends analyses; this GC sample consists of 25 red and 14 blue GC sources. We also defined a high significance sample, including only GC sources whose (0.3-8.0 keV) flux was measured at $\geq 3.0\sigma$ confidence (see B09), consisting of 21 red and 14 blue GC sources. The sample of field sources used for statistical work includes 71 sources, of which 55 are in the $\geq 3\sigma$ sample.

## 3. Spectral Analysis of Individual Bright Sources

We used the CXC CIAO software suite (v4.0.1) and CALDB version 3.5 to extract the spectral data and associated calibrations for each of the six observations individually to produce separate spectral (PI), ancillary response (ARF) and redistribution matrix (RMF) files for each (as described in http://cxc.harvard.edu/ciao/ahelp/psextract.html).

To extract spectral data, we used circular regions centered on each source, with radii chosen so as to ensure that as many source photons as possible were included, while minimising contamination from nearby sources in the crowded source field of NGC 4278 (B09). Background counts were extracted for each source individually from an annulus surrounding the source, with inner radius equal to the source extraction radius, and outer radius at least three times that value. These source extraction regions in some cases may be smaller (with a minimum radius of 1.5") than those in B09, where circles of 3.0" radius were used and pie segments containing contaminating sources excluded (and then area corrections applied). Given the Chandra PSF and the small off-axis angles of our sources, these smaller radii would imply a maximum loss of $<$7% of source counts, which will not affect our results. We investigated the effect of using smaller circular extraction regions in the results of the spectral analysis using the relatively isolated luminous source S96 (for which the extraction radius was 2"), and no significant differences were found from the smaller extraction regions spectra. In particular, the increase of source counts is of 3% (25 counts) and the spectral results are not affected; although some of these counts would be found in the background extraction region, only 1-2% of the source counts are from the background, so the different amounts of background photons would not be a concern, even if they were not subtracted. Any detected sources that lay within the background annulus were excluded using circular cut-outs and the outer annuli radii were extended to account for the lost extraction area. Given the typically very low and non-varying background of the *Chandra* ACIS data, the use of large background regions is not a problem. Fig. 1 identifies the luminous sources selected for individual analysis in the joint *Chandra – HST* field of view of NGC 4278 and Fig. 2 shows examples of source and background source extraction areas.

We used observation specific bad pixel files to account for the decrease in the efficiency of the detector due to bad pixels and cosmic ray flux disabled pixels. The spectral data were binned in groups of at least 20 counts to allow for Gaussian errors when using $\chi^2$ statistic. No binning was used for Cash statistic (Cash 1979), which we used for analyzing low-count-rate spectra. Recent work investigating parameter biases from maximum-likelihood statistics, by means of Monte Carlo simulations, has suggested that the Cash statistic should be used in preference to the $\chi^2$ when modeling Poisson-distributed data, even in the high count regime, where $\chi^2$ can produce 5 – 10% biased model parameter estimates (Humphrey et al. 2009). In our case, by comparing the Cash and $\chi^2$ statistics in both high count spectra with $\sim$ 1000 counts and spectra with only $\sim$ 200 counts, we have determined that both statistics provide the same best-fit parameters, within errors, indicating that our results do not appear to be affected by the choice of statistic. The spectra were analyzed with XSPEC (v 11.3.2ag) using an energy range of 0.3-8.0 keV to fit either single component power-law (PO) or accretion disk (DBB) models along with the *wabs* photoelectric absorption model (as discussed in B10, given the statistics of our data and the ACIS spectral resolution, our results are indifferent to the use of either *wabs* or *phabs* absorption model). We chose these models, rather than the bremsstrahlung spectrum used in some other studies (e.g. Kim & Fabbiano 2003, Sivakoff et al 2008), because PO and DBB spectra have been consistently used as baseline models for fitting the X-ray spectra of Galactic LMXBs (see e.g. McClintock & Remillard 2006). Using these models, therefore, will allow us to compare directly the spectra of the NGC 4278 LMXBs with those of their well-studied Galactic analogues.

The data of each source were grouped for fitting based on the detected luminosities and hardness ratios (HR) from B09 (Figs. 3, and 4), as discussed below. In joint fits of different data sets, all model parameters were linked in XSPEC, except for the normalization. We first performed fits letting the HI column density $N_H$ free to vary, and 2-parameter confidence contours were derived using these results, even if the best-fit $N_H$ was lower than the Galactic value along the line of sight $N_H(Gal.) = 1.76 \times 10^{20}$ cm$^{-2}$ (calculated with the tool COLDEN: http://cxc.harvard.edu/toolkit/colden.jsp; COLDEN uses the data of Stark et al 1992 and Dickey & Lockman 1990). These confidence contours were used together with the simulations of B10 to constrain the spectral models (see discussions below; also B10). However, if the best-fit value was smaller than the line of sight $N_H(Gal.)$, $N_H$ was then fixed at $N_H(Gal.)$ and the fit repeated to calculate the second best-fit parameter and its uncertainties, reported in the Tables. As discussed below, in these cases the adopted spectral model is not likely to be the correct one, so strictly speaking these results are only of interest for comparison with previous results in the literature and they are only reported for completeness. We find that these unphysical small $N_H$ values tend to occur in DBB fits and are a symptom of the presence of a power-law component in the spectrum (B10). Fig. 5 shows examples of source spectrum, best fit model and residuals from our analysis.

To investigate the effect on the fit parameters of fitting a spectrum including different power-law (or DBB) component, such as it may occur when considering several spectra of a given source or the spectra of several sources together (see below and § 4), we have used a similar simulation approach as in B10. However, for these simulations instead of using a two-compontent DBB+PO model, we have used two PO *or* DBB models. In the case of the PO composite model we fixed the first component to have $\Gamma$=1.7 and varied the photon index of the second component to include slopes of 1.2, 1.5, 1.7, 2.1 and 2.3. The first component of the input DBB model was defined to have an inner disc temperature of 1.00 keV and the second component had kT values of 0.50, 1.00, 1.50 and 2.25 keV. In each set of simulations the normalizations of both components were adjusted to provide values of 90%, 75%, 50% and 40% flux ratio arising from the second PO (or DBB) component. As in B10, each model was defined to generate a spectrum of 1000 counts and 100 simulations were produced for each set of parameters. Each of these simulations were then modeled by a single-component PO and DBB model with the *phabs* photoelectric absorption model.

The results are shown in Table 2, and indicate that the resulting power-law index (or DBB kT) is in-between those of the input parameters, so the result gives an indication of the average spectrum. As in B10, when modeling a thermal spectrum with a PO model the value of $N_H$ is greatly elevated compared to the Galactic absorption value. Further, when applying a DBB model to the non-thermal spectrum the value of $N_H$ is 0 in all cases.

These simulations (alongside the results of B10) indicate that when fitting the spectra of several sources together, the dominance of both thermally dominated and hard states can be inferred from the $N_H$ value of the fit. If the absorption column values indicate that the source is in a non-thermal or thermal state, the value of $\Gamma$ (or kT) can be taken as an average of the source spectra included in the fit.

### 3.1. Luminous GC LMXBs

The GC sources S96, S163, S185, and S194 of B09 were selected for individual analysis, based on the luminosity criterion discussed in §2. When using a 'canonical' power-law model ($\Gamma = 1.7$; $N_H = N_H(Gal.)$, see B09), S96 was detected with co-added average luminosity of $\sim 3.8 \times 10^{38}$erg s$^{-1}$ and the other three sources with similar coadded luminosities of $\sim 2.5 \times 10^{38}$erg s$^{-1}$ (Fig. 3). S96 and S194 are associated with blue GCs, S163 and S185 with red GCs.

The results of the spectral analysis are summarized in Table 3, which lists: source number; observations used (following Table 1); net source counts and count rate, both with $1\sigma$ statistical error; model; Cash statistics; reduced $\chi^2$, number of degrees of freedom and probability (when using Cash statistic, we present a value of 'goodness' in place of probability. This value represents the percentage of Monte Carlo simulations produced from the best-fit data model, which return a lower value of Cash than has been obtained with the source data); best-fit $N_H$, $\Gamma$ and kT, all with $1\sigma$ error for one interesting parameter (F indicates that $N_H$ was frozen at the Galactic value); (0.3-8.0 keV) absorption-corrected best-fit luminosity, calculated from exposure-weighted luminosities for variable sources. B09 shows that S96 is fairly steady in both luminosity and HR throughout the six observations, S185 experiences significant changes in flux across the different observations with no HR variation, while S163 and S194 change in both luminosity and HR multiple times (see Fig. 3). We show the results for the fits of different portions of the data (see below) and of the 'grouped' spectra of each source in Table 3. Fig. 6 shows 1 and $2\sigma$ confidence contours for two interesting parameters for both PO ($N_H$ and Γ) and DBB ($N_H$ and kT) models, for the different spectral/luminosity states of S163 and S194. The confidence contours for the coadded spectra of the four sources are compared in Fig. 7. In all cases, there is no strong statistical preference for a PO or DBB model, but the allowed parameter spaces for a given model may differ significantly in some cases. The above applies to all the results presented in this papers.

For S163 we grouped the data as follows: Group 1 contains the softest and highest count rate observations ($L_X \sim 4 \times 10^{38}$ erg s$^{-1}$ in Fig. 3, where the luminosity is from B09, calculated for the same assumed spectrum with $N_H = N_H(Gal)$ and $\Gamma = 1.7$ in all cases, therefore it represents just a scaling of the detected count rate) and Group 2 contains all the other observations that have similar, harder spectra compared to Group 1. We also subdivided the Group 2 observations based on luminosity: Group 3 contains the two intermediate count rate observations, with B09 $L_X \sim 3 \times 10^{38}$ erg s$^{-1}$, and Group 4 contains the two lowest count rate observations, with B09 $L_X \sim 1 \times 10^{38}$ erg s$^{-1}$. Although the parameter spaces overlap, the best-fit values suggest that S163 alternates between a high/soft state, with $\Gamma \sim 2.5 \pm 0.5$ (or kT$\sim 0.7 \pm 0.2$keV) and a low/hard state with $\Gamma \sim 1.4$ (or kT$\sim$ 1.6 keV), before finally settling down in the low/hard state. With the exception of the PO fit of the high/soft Group 1 spectrum, which requires intrinsic absorption with $N_H \sim 2 \times 10^{21}$ cm$^{-2}$, $N_H$ is consistent with $N_H(Gal.)$. The larger $N_H$ in the power-law fit of the high luminosity data suggests spectral curvature, favoring the DBB model (see Fig. 8-Left, based on the simulations of Brassington et al 2010 - hereafter B10).

A similar procedure was followed for S194: Group 1 contains the two highest count rate observations; Group 2 contains a single observation which has markedly low HR and intermediate count rate; Group 3 contains the observations with the lowest count rate, and with HR consistent with that of Group 1. S194 appears to go through two distinct state transitions, beginning with a high/hard state with a B09 luminosity of $\sim 3 \times 10^{38}$ erg s$^{-1}$ and $\Gamma \sim 1.8 \pm 0.4$ (or kT$\sim 1.1 \pm 0.1$ keV), changing to a high/soft state with a slightly lower B09 luminosity of $\sim 2 \times 10^{38}$ erg s$^{-1}$ and slightly softer spectrum (kT$\sim 0.7 \pm 0.1$ keV), and then settling in a luminosity-independent hard state. In the power-law fit, the highest luminosity Group 1 has $N_H > N_H(Gal.)$ while $N_H$ is consistent with $N_H(Gal.)$ in the DBB fit, which may be indicative of a thermal disk dominant emission (see B10). The other spectra are all consistent with $N_H(Gal)$ in the PO fits, which may be indicative of a significant power-law component (see Fig. 8-Right, based on B10).

The grouped spectra of all the observations for each of the four sources, except S185, are consistent with $N_H(Gal.)$ for both models. These spectra are fitted with 'average' $\Gamma \sim 1.3 - 1.7$ in the PO model; the best fit PO for S185 instead has $N_H$ larger than line of sight. The latter result may be indicative of a significant disk component (see Fig. 8-Left). The DBB fits return generally consistent values of $kT \sim 1.2 - 1.5$ keV. Using the best fit models for the coadded spectra, we derive luminosities slightly higher than the values in B09, but the differences are understandable given the differences of the best-fit parameters with those assumed in B09 (power-law spectrum with $N_H = N_H(Gal)$ and $\Gamma = 1.7$). In all cases, the luminosity variability patterns of B09 are still present.

### 3.2. Luminous Field LMXBs

The results of the spectral analysis of three B09 field sources, S146, S158 and S184, each with average B09 luminosity of $\sim 5.0 \times 10^{38}$erg s$^{-1}$ over the six observations (Fig. 4), are summarized in Table 4, which follows the same format as Table 3. The contours for the two interesting parameters for both PO and DBB models for the three sources can be seen in Fig. 9.

For S146, Fig. 4 shows that the HR does not vary; the B09 luminosity (count rate) is also constant, except for Obs. 2, where it is a factor of 6 lower. We divided the data in two groups: Group 1, which excludes Obs. 2; and Group 2 (Obs. 2). The PO fits for both groups give $N_H \sim 10 \times N_H(Gal.)$; $\Gamma$ is $1.4 \pm 0.2$ for Group 1 and larger ($\Gamma = 2.1 \pm 0.3$) for Group 2 (see Fig 9). The DBB fit gives $N_H$ consistent with $N_H(Gal.)$; reflecting the steeper spectrum, the lower-luminosity Group 2 has a lower value of kT (1.0$\pm$0.2 keV against 1.6$\pm$0.2 keV for Group 1). The $N_H > N_H(Gal)$ in the PO fits suggests that the DBB model may be a better representation of the emission spectrum, or that a sizeable disk component may be present in the emission (see simulations of B10; Fig. 8-Left).

For S158, Fig. 4 shows that there are variations in both luminosity and HR between different observations. Accordingly, we fitted jointly spectra as follows: Group 1 contains the observations which show the hardest spectra, and have a similar B09 luminosity of $\sim 5 \times 10^{38}$ erg s$^{-1}$; Group 2 contains the observation with the softest spectrum, which also has a B09 luminosity in the range of Group 1; Group 3 contains the observations which have the lowest B09 luminosity ($\sim 3 \times 10^{38}$ erg s$^{-1}$) and intermediate HR. The results echo those for S146, with the PO model requiring intrinsic absorption, while the DBB model gives results consistent with the line of sight $N_H$ (but best-fit $N_H > N_H(Gal.)$, except for Group 3).The confidence contours show that the values of $\Gamma$ (or kT in the DBB model) are consistent within $2\sigma$, although the Group 2 parameter range suggests a softer spectrum. As for S146, the $N_H > N_H(Gal)$ in the PO fits suggests that the DBB model may be a better representation of the emission spectrum (see simulations of B10; Fig. 8-Left).

For S184, both luminosity and HR variations are small. Group 1 contains observations 1 and 6, which displayed the highest B09 luminosity of $\sim 5 \times 10^{38}$ erg s$^{-1}$ and consistent HR; Group 2 contains observations which had a slightly lower B09 luminosity of $\sim 4.5 \times 10^{38}$ erg s$^{-1}$, and also have similar HR (Obs. 2, 3, 4); Group 3 contains Obs. 5, which has consistent $L_X$ with Group 1, but different X-ray colors. The PO model fits give larger $N_H > N_H(Gal.)$ for the higher luminosity

Group 1 and Group 3, together with a steeper power-law index, when compared to the Group 2 results (Table 4, Fig. 9). This again may suggest a larger disk fraction in the spectrum of Groups 1 and 3 (see B10). The DBB fits are in both groups consistent with $N_H(Gal.)$ and also have similar kT$\sim$ 1.5 – 1.6 keV.

The grouped total spectra (Fig. 10) of the three sources in the PO fits have best-fit $N_H$ exceeding $N_H(Gal.)$, although the contours are marginally consistent with this value, except for S158, which shows definite intrinsic absorption. The DBB parameter spaces are instead consistent with $N_H(Gal.)$ in all cases. S158 (in both models) appears marginally softer. We derive best-fit luminosities slightly exceeding those in B09. The luminosity variability patterns of B09 are reproduced, with the exception of S158. In the case of S158, the variability behavior derived from B09 differs to that inferred here. This discrepancy arises from the different extraction regions used in the source counts photometry. In B09, S158 has been classified as an overlapping sources and therefore a pie-sector correction was made to account for this (see §2.1 of B09 for details). Here a smaller extraction region was used to exclude any counts from the overlapping source. For both methods this resulted in a reduction in source count of $\sim$5% for observations 1,2,3 & 4. However, for observations 5 & 6 there was a reduction of $\sim$50% in B09, compared to only 7% & 9% in this work. From inspection of the raw counts this is a consequence of the asymetric distribution of source counts in observations 5 & 6.

## 4. Spectral Analysis of Field and GC LMXB Samples: Luminosity and Colors

To extract data for the spectral analysis of the source samples defined in § 2, we followed the same process as in § 3, except for S174, where we used an elliptical extraction region, to avoid overlap with an adjacent source. The background was extracted from a large annulus centered on the galaxy, with inner radius that extended beyond the densely populated central region and an outer radius that reached the $D_{25}$ ellipse. The source and background region files were then processed using the *specextract* command in CIAO, creating a set of weighted ARF and RMF files, which take into account the spatial as well as the temporal variation of the instrument properties. The weighting was done using the raw counts to produce a map of the distribution of counts across the image regions. Bad pixel and dead area correction were both applied as in the single source case, and binning groups were set to a minimum of 20 counts each for the $\chi^2$ fits. Each set of spectra was then fitted with PO, DBB and composite PO+DBB models in XSPEC, with the PI files from each observation being fitted jointly for each luminosity/colour group. The results of the model fits for these samples are summarized in Tables 5, 6 and 7, which list: the sample used, net source counts and 1$\sigma$ statistical error, the model used for the fit, the number of degrees of freedom for the $\chi^2$ fit, the corresponding probability, the best-fit $N_H$ with 1$\sigma$ error for one interesting parameter and the best fit Γ and kT with $1\sigma$ error (F denotes $N_H$ frozen to $N_H(Gal.)$).

While background contamination cannot bias the results of the analysis of the individual luminous sources, because of the narrow PSF and low background of *Chandra* (see §3), it may in principle become an issue here, given the large composite extraction areas of the joint spectra. The high and mid $L_X$ sample results would not be affected, because 97% of the counts in the combined extraction regions arise from the sources, therefore, given the dominance of the source counts, eventual diffuse emission is not going to alter the fit. The issue does become pertinent when we consider the low $L_X$ samples, where 30%-40% of the extracted counts are due to background. However, only one of the low-luminosity GC sources and five of the field sources lay in the region potentially most affected by excess diffuse emission (i.e., hot ISM in NGC 4278), within the inner radius of the background annulus (20”), with no sources within the central 10", where the diffuse emission is most concentrated (Trinchieri et al 2010, in preparation). It is therefore likely that there is little contamination from a diffuse emission component in the total spectrum. Nevertheless, we tried including an additional APEC thermal component in the fits, to account for any potentially-unsubtracted diffuse emission, and found no improvement in the fit-statistic; moreover, the best-fit APEC kT were close to 0 keV in all runs, indicating that this additional component is not required.

PO + DBB fits were also attempted, with $N_H = N_H(Gal.)$; although the best-fit Γ and kT were generally found to be consistent with those of the single-component fits, the fit parameters, including the normalizations, were unconstrained (as determined by inspection of 2-parameter confidence contours). We therefore do not report these results below.

### 4.1. GC LMXBs

Table 5 gives our results for PO, DBB and combined PO+DBB model fits for the high, mid, and low luminosity GC source groups. Fig. 11(top) shows the 2-parameter 1 and $2\sigma$ confidence contours for the PO and DBB models. For the PO model, the mid and high-luminosity spectra have consistent parameter spaces, with best-fit $\Gamma = 1.5$, and best-fit $N_H$ exceeding the Galactic $N_H$, although the parameter space of the mid-luminosity sample is consistent with this value. The low-luminosity group parameter space suggests an intrinsically absorbed softer spectrum with Γ $\sim 2.0\pm0.15$, which is inconsistent with the higher luminosity groups at the $> 2\sigma$ confidence level. The elevated best-fit $N_H$ required by the PO fits may point to a dominant average thermal emission (see Fig. 8-Left, B10). The DBB best-fit $N_H$ is below the Galactic $N_H$ for all the spectra, however the Galactic value is within the allowed parameter space in all cases; at $N_H = N_H(Gal.)$, the mid and high luminosity spectra are fitted with kT in the range $\sim 1.3-1.6$ keV, while the low luminosity spectrum is consistent with a lower kT of $\sim 1.1\pm0.09$ keV, indicating a softer spectrum, as in the case of the PO fits.

Since the high and mid luminosity confidence spaces are consistent, we also combined these data for a single fit, in order to improve the statistics of the results. The resulting $N_H$ significantly exceeds $N_H(Gal.)$, and $\Gamma$ remains near 1.5. The combined parameter space is inconsistent with that of the low luminosity group, at $> 3\sigma$ confidence (not shown). For the DBB fit to the combined high+mid spectra, freezing $N_H = N_H(Gal.)$ gives a best fit temperature of 1.5keV. Overall the fits are good, with most reduced $\chi^2_\nu$ values around 1.0, the exception being the high, and high+mid luminosity samples which have $\chi^2_\nu \sim 1.4$. Using PO+DBB models does not improve the quality of the fit (not shown). One possible reason for the poorer fits of the high luminosity spectrum is the small number of sources included, with variable spectra, which may not be approximated adequately by the simple fit models. There are only 4 high luminosity GC sources, two of which show long term variability and two that show short term variability (B09).

### 4.2. Field LMXBs

Table 6 summarizes the results for PO and DBB and combined PO+DBB model fits for the high, mid and low luminosity field LMXB samples. The large difference between the sum of the 'high', 'med' and 'low' counts and the counts in the 'all' group is a consequence of the omission of some field sources in the three luminosity groups, which have been included in the 'all' group. As discussed in § 2, these sources were excluded from the sub-groups due to large overlaps with other field sources with different luminosities cuts.

The 2-parameter confidence contours for the PO and DBB fits are shown in Fig. 11 (bottom). The $3\sigma$ 2-parameter power-law contours of the high luminosity field spectrum (not shown,) are inconsistent with either the medium or low luminosity spectra, requiring intrinsic absorption in excess of $N_H(Gal.)$; as discussed in §3, this result may suggest a dominant disk emission in these luminous sources (see B10, Fig. 8-Left). The medium and low-luminosity parameter spaces overlap at $2\sigma$ confidence values, and are consistent with $N_H = N_H(Gal)$, suggesting an average power-law dominated spectrum (see B10). Fixing $N_H$ for these two sub-samples at the Galactic value, the allowed intervals of photon indices are inconsistent at $> 2\sigma$. All the spectra are consistent with $\Gamma \sim 1.6 - 1.7$, however the medium-luminosity spectrum may be consistent with a steeper power-law ($\Gamma \sim 1.8 \pm 0.08$) when $N_H$ is frozen at Galactic value.

In all cases, the DBB fits tend to have minimum $\chi^2$ at $N_H < N_H(Gal.)$; moreover, the confidence contours for medium and low luminosity sample exclude $N_H(Gal.)$, suggesting the presence of a low-energy excess in the data, which may be connected with the presence of a significant power-law component (see Fig. 8-Right, B10). Freezing $N_H = N_H(Gal.)$, we obtain best-fit kT$\sim$1.4, 0.9 and 1.2 keV for high, medium, and low-luminosity spectra respectively, with $\chi^2_\nu$ values of 1.2, 1.6 and 1.1, respectively (see Table 6). So while the DBB fits are acceptable in this instance for both high and low-luminosity spectra, the PO model provides a significantly better fit for the medium-luminosity spectrum. As shown in the Table 6, the values of kT found are inconsistent with one another at the $2\sigma$ level, if we freeze $N_H = N_H(Gal.)$. However, as discussed above, our comparison with the B10 simulations suggests that DBB only emission is not likely, especially in the lower luminosity samples.

### 4.3. Comparison of Sample Spectra

The results of the joint fit of coadded spectra in selected luminosity ranges show differences when comparing GC and Field samples. The high luminosity contours (solid in fig. 11) do not overlap in the power-law fit, but the range of $\Gamma$ is the same, while the field sources are fitted with significantly higher $N_H$.; the contour are instead consistent in the DBB fit. The mid-luminosity contours (dotted in fig. 11), have a similar range of $N_H$ in the PO fit, but the range of $\Gamma$ is steeper for the field sources (1.7-2 versus 1.3-1.7 for the GC spectra). This difference is reflected in the DBB contours, where the field temperatures are lower; however the latter spectra are also badly fit with the DBB model, as discussed above, and require in this model unphysical low $N_H$, suggestive of a power-law spectrum (B10, fig. 8-Right).

The low-luminosity parameter spaces (dashed in Fig. 11) only partially overlap in the PO fits, with the GC spectrum being fitted with steeper power-laws and requiring $N_H$ higher than Galactic, while the field spectrum is consistent with Galactic $N_H$. The DBB parameter spaces have consistent kT$\sim$ 1.1 keV, but require best-fit $N_H < N_H(Gal.)$ for the field LMXB spectrum, consistent with a power-law component in the emission (B10, Fig. 8-Right); although the best-fit $N_H$ is also below the Galactic value for the GC spectrum, $N_H(Gal.)$ is within the confidence contours.

Table 7 summarizes the results of the spectral fits for the sources associated with either red or blue GCs. Fig. 12 shows the two-parameter confidence contours for the PO and DBB model fits for these spectra. For the PO model, the $1\sigma$ parameter spaces barely touch, suggesting slightly lower $N_H$ in the red GC spectra; however, the $2\sigma$ parameter spaces are consistent. For both red and blue samples the PO fit is consistent with $N_H > N_H(Gal)$. In the DBB fits, the red GC parameter space is within that of the blue sample, and again suggests lower $N_H$. Although, both red and blue GC spectra have unphysical low best-fit $N_H$ in the DBB fits, only for the blue sample is the allowed parameter space consistent with $N_H(Gal.)$.This result is consistent with the red sample having a predominantly power-law spectrum (see B10), the power-law $\chi^2$ is also more acceptable for the red spectrum than the DBB $\chi^2$.

## 5. Trends and Correlations

We have analyzed the NGC 4278 GC samples defined in §2, searching for possible relations between the presence of X-ray sources in GCs and the X-ray luminosity of these sources versus the GC parameters: $V$ magnitude, $V-I$ color, and galactocentric radius $R_G$. $V$ is directly related to the stellar content and mass of the GC, $V-I$ has been related to both age and metallicity of the GC stellar population (see review, Brodie & Strader 2006), $R_G$ gives the position of the cluster relative to the center of the parent galaxy, and could then be related to possible dynamical effect on the GC due to its interactions with the associated galaxy.

We confirm previous reports (see Angelini et al 2001, Kundu et al 2002, review in Fabbiano 2006) that more luminous GCs have a higher probability of hosting an X-ray source. Table 8 gives the number of GCs (red and blue) in two magnitude ranges, the number of associated LMXBs, and the fraction of GCs associated with an LMXB. This table shows that - no matter the $V-I$ color of the GC - , LMXBs are more likely to occur in the minority of GCs with $V \leq 22.5$ mag. In the entire GC sample, 50% of GCs with $V < 22.5$ host an X-ray source, against 10% of GCs with $V > 22.5$; the median value for the V band magnitude of all GCs in NGC 4278 is 23.1, whereas the median value for those with an X-ray source is 22 mag. This trend is evident in Fig. 13. The same figure shows no indication of an $R_G$ dependence for the probability of a GC hosting an LMXB, in agreement with the results of the Virgo survey of Sivakoff et al (2007). We also confirm the prevalent association of LMXBs with red, rather than blue, GCs, which has been widely reported in the literature (see Fabbiano 2006 and refs. therein). Table 7 shows that out of a total of 121 red and 145 blue GCs in NGC 4278, X-ray sources have been identified with 25 red clusters, and 14 blue, representing 21% and 10% of their respective populations.

With our sample we cannot confirm the association of higher $L_X$ LMXBs with red GCs (e.g. Kundu et al 2007). Fig. 14 (top right) shows no obvious $L_X-(V-I)$ trend; Table 9 shows that the fractions of sources in the high, mid and low $L_X$ groups are consistent (within statistics) for the red and blue GC-LMXB samples. Spearman Rank (SR) tests give probabilities of chance correlation $P = 27.3\%$ and 38.3% for the full and $\geq 3\sigma$ samples respectively, excluding an $L_X-(V-I)$ dependence in our data. Posson-Brown et al (2009) similarly report no $L_X-(V-I)$ trend in NGC 4636.

Fig. 14 (left) shows scatter diagrams of $L_X$versus $V$ and $R_G$ for sources associated with GCs. While a weak trend may be present in the $L_X-V$ diagram, a more definite correlation is suggested by the $L_X-R_G$ diagram, which shows a definite lack of high $L_X$ sources at large galactocentric radii in the GC-LMXB sample of NGC 4278. This trend is not present in the field LMXB sample (fig. 14, bottom right). To explore the possibility of correlations, we performed SR tests on both samples. To minimize the effects of source confusion, we ignored all the sources with $R_G < 10$”; this cut affects only the field LMXB sample. Moreover, we performed tests both including and excluding sources with flux determined at $< 3\sigma$ confidence; the $3\sigma$ flux cut effectively excludes all sources fainter than $\sim 1\times10^{37}$ erg$s^{-1}$. As can be seen from fig. 14, these fainter $< 3\sigma$ sources are uniformly distributed across $R_G$, so that their exclusion would not bias our results. The results of the SR tests are summarized in Table 10. The SR test yields a very small probability of chance correlation $P = 0.084\%$ for the sample of 39 GC LMXBs (0.038% for the sample of $35 \geq 3\sigma$ sources), confirming a significant $L_X - R_G$ correlation; the SR chance probability is instead $P =$ 4.5% for the sample of 58 field sources ($P = 11.3\%$ for the $43 \geq 3\sigma$ field sources), excluding a significant correlation for field LMXBs.

We can exclude that the GC-LMXB $L_X - R_G$ correlation is a second-order effect of stronger correlations with a third parameter, the GC optical magnitude $V$, which may arise if more luminous GCs are found at smaller galactocentric radii (e.g., Goudfrooij et al 2004, 2007). In the entire WFP2 GC sample of NGC 4278 there is no $V - R_G$ correlation, at best a weak trend (Fig. 13); SR chance probabilities are 6.1 and 14.7 % for the red and blue GC samples respectively, and 5.5% for the combined GC sample. Moreover, we find no evidence of a $L_X - V$ correlation (Fig. 14); SR $P = 11\%$ for the full sample, and 27% for the $\geq 3\sigma$ sample.

Concluding, our analysis of correlations between $L_X$, optical magnitude $V$, and galactocentric distance $R_G$ (Table 10) strongly suggests a $L_X - R_G$ correlation in the GC-LMXB sample of NGC 4278, which is not caused by a GC luminosity effect.

## 6. Discussion

### 6.1. Luminous GC and field sources - Similarity with Galactic LMXBs

Because of their luminosities in the $\sim 3-8\times10^{38}$erg s$^{-1}$ range (B09, Tables 2 and 3), the seven sources that belong to this category are either NS binaries emitting above the Eddington limit, or BH binaries (BHB). The intensity/spectral variability results provide further insights on the nature of some of these systems. Galactic BH LMXBs typically show X-ray spectra that can be fitted with either a power-law, or a multi-color disk model, or a combination of these two models. These sources typically alternate between spectral states, which can be either accretion disk dominated (the thermal or high/soft state) or power-law dominated (the low-luminosity hard state, and the high-luminosity steep power-law or very high state; see reviews: McClintock & Remillard 2006; Remillard & McClintock 2006; Done et al 2007). These power-law states have been modeled in terms of varying accretion rates, resulting in either Comptonized emission and/or jets. Similar spectral behavior has been observed in Galactic NS and BH LMXBs, the determining difference being the presence of a hard boundary-layer component in NS sources, which may arise from the NS surface - disk interface. (e.g., Done et al 2007; Fender et al 2004).

Our data do not have enough statistics to allow meaningful results for a composite power-law plus disk model, but can be fitted with both individual power-law (PO) and disk (DBB) models (§3). When fitted with disk models, the average spectrum of each source is consistent with best-fit $kT_{in}$ in the range $\sim 0.7-1.7$ keV; the power-law fits give a range of best-fit $\Gamma \sim 1.2-2.0$. Overall, except for the low end of the power-law indeces, these values are in the range of those found in BHBs (see McClintock & Remillard 2006). The flatter power-laws could be indicative of the increased 1-10 keV emission due to the boundary layer in NS binaries (see fig. 26 of Done et al 2007), although the luminosities of these sources exceed the NS Eddington luminosity.

Of the four GC LMXBs in this group of sources, two (S163 and S194) appear to undergo luminosity/spectral transitions (see Fig. 3). S163 may alternate between a higher count-rate disk-dominated state and a lower count-rate hard state. Although the changes in count rate are of a factor of $\sim 4$ at the most, there are Galactic BH binaries, showing this type of spectral/count-rate transitions (e.g. H1743-322, XTE J1859+226 and GX339-4; see Remillard & McClintock 2006). As discussed in §3.1, the the high luminosity state is characterized by a disk temperature of $kT_{in} = 0.7 \pm 0.2$ keV when the data are fitted with a single DBB model; this temperature is in the range of temperatures measured in BHBs in thermal state (see McClintock & Remillard 2006). The single power-law fit returns a relatively steep index ($\Gamma \sim 1.5-2.5$) and requires significant intrinsic absorption; the simulations of B10 demonstrate that a spectrum consisting of both power-law and significant disk emission would give rise to spuriously high best-fit $N_H$, when fitted with a single power-law model. All of the above suggests that we have observed S163 in a thermally-dominated state.The lower count-rate observations instead are fitted with power-law models with $\Gamma \sim 1.4$, and absorption consistent with Galactic $N_H$. These power-laws may suggest a 'hard' perhaps jet dominated state, although the indeces are in the low range of those measured in Galactic BHBs in hard state (Remillard & McClintock 2006; McClintock & Remillard 2006). We may rule out a simple thermal disk dominated emission in this source because the single DBB fits return higher disk temperature (kT$\sim 1.6-2.0$keV) than that measured at higher count rate, resulting in a $L_{disk} - T_{in}$ relation much shallower than the $L_X \sim T_{in}^4$ relation expected in disk-dominated systems (e.g., Makishima et al 2000); however, this result may derive from the presence of Comptonized radiation, as in the case of GRO J1655-40 (Kubota et al 2001).

In S194, except for the highest count-rate state where $N_H > N_H(Gal.)$ in the power-law fit suggesting a thermal dominated state, no strong $N_H$ signature is detected. If these spectra are thermal, they may suggest fluctuations of the disk temperature, in response to fluctuations in the accretion rate (see e.g. Smith et al 2002; Fabbiano et al 2003). The results are consistent with the $L_X - T_{in}^4$ relation.

Changes in luminosity do not necessarily result in a change in the spectral state: the GC source S185 changes in count rate by a factor of $\sim$2, however the spectral parameters are steady, a behavior observed in some Galactic BH candidates (McClintock & Remillard 2006). However, this source also has the smallest number of counts of any of the individual analyzed sources, and so it is possible that changes in spectral state went undetected.

The luminous field sources S146, S158 and S184 may all largely be in a thermal disk-dominated state, since in most cases the power-law fits result in large intrinsic $N_H$ values (see §3.2). Given the substantially super-Eddigton luminosities of these sources ($> 5 \times 10^{38}$erg s$^{-1}$) for a NS counterpart, it is likely that they are BHBs. In S146 we may have observed the result of a drop in accretion rate during the second *Chandra* observation, resulting in a drop in both luminosity and disk temperature. For S158, our results suggest that this source is first observed in a high/soft state, becomes softer with decreasing luminosity possibly because of cooling of the disk and then may transition to a low/hard state with an increasing larger power-law component. This last transition is suggested by the power-law contours moving towards the galactic $N_H$ line, and similarly the DBB contours allowing values of $N_H < N_H(Gal.)$ (Fig. 11), suggesting excess low energy emission above the DBB model that could be due to a prominent power-law component (see Fig. 8, B10). For S184, the differences between the spectra occur only in the power-law fits, while the confidence contours of the DBB fits overlap and allow Galactic line of sight $N_H$, although the best-fit $N_H$ of the lower luminosity spectra falls well below the Galactic value. Therefore, while we do not find a strong signature for a state transition, the spectral results may suggest that the source is transitioning between an initial high/soft state and a low/hard state.

Fig. 15 displays the results of the spectral fits in $L_X - T_{in}$ (or $L_X - \Gamma$) plots, using the additional constraints from the B10 simulations (see Section §3). Whenever it could not be established that a given spectrum was either disk or power-law dominated, the results were plotted in both diagrams. The analogous results of B10 for the luminous LMXBs of NGC3379 are also plotted. The thermal spectra are generally consistent with increases of disk temperature with increasing luminosity, and within the uncertainties with the $L_X - T_{in}^4$ relation. These results suggest that these luminous LMXBs, both in the field and in GCs, are consistent with accreting BHBs with BH masses in the $\sim 5-15$ solar mass range. Interestingly, this is the range of masses found in Galactic BHBs (see Ozel et al 2010), further reinforcing the connection between these extragalactic LMXB populations and their Galactic counterparts.

Some of these sources are in GCs.The possibility of BHBs forming and residing in GCs is a debated issue, since none has been found in the Milky Way, and theoretical expectations regarding their properties, transient behavior, and detectability argue that 10 solar mass BHBs should be hard to discover compared with NS binaries (Kalogera et al 2004; see Ivanova et al 2010 for a recent discussion of BH binary formation in GCs). However, very luminous GC sources have been found with *Chandra* in several early-type galaxies, suggesting that GC BHBs exist (see Verbunt & Lewin 2006; Kim E. et al 2006). In one case at least, in NGC 4472, both the high luminosity ($\sim 4 \times 10^{39}$erg s$^{-1}$) and a factor of 7 intensity variability provide a convincing candidate of a GC BHB (Maccarone et al 2007). In the elliptical galaxy NGC 1399, the optical spectrum of a GC hosting a luminous X-ray source may suggest the presence of a $1000 M_{\odot}$ BH (Irwin et al 2010). Our work on the luminous LMXBs in NGC 3379 has also suggested GC 'stellar mass' BHB candidates, and possibly a more massive BH (B10).

### 6.2. Average spectral properties of LMXB samples

The analysis of the joined spectra of the three luminosity samples for the field LMXBs (§4) suggests that the average spectrum of the four high-luminosity sources is more likely to be thermal, while the average spectra of medium ($6 \times 10^{37} – 1.5 \times 10^{38}$ erg s$^{-1}$) and low-luminosity ($L_X < 6 \times 10^{37}$ erg s$^{-1}$) sources may include strong power-law components. The high luminosity joined spectrum is strongly inconsistent with the line of sight Galactic $N_H$, requiring considerable intrinsic absorption in the PO fit; the DBB fit, however, is consistent with $N_H(Gal)$. Looking at the spectral simulations of B10, this result suggests the presence of a significant (or dominant) thermal DBB component in the emission (Fig. 8-Left). Conversely, the mid and low-luminosity results require $N_H$ significantly below $N_H(Gal.)$ in the DBB fit, suggesting the presence of a power-law spectrum (Fig. 8-Left-Right; see B10). Smaller $N_H$ values in lower-luminosity LMXBs are also reported in the spectral study of the NGC 4697 LMXB population (Sivakoff et al 2008); these authors however do not comment on possible fit biases and suggest 'real' accumulation of absorbing material in these sources. While we cannot exclude this hypothesis, we believe that a firm conclusion on intrinsic luminosity-dependent $N_H$ cannot be supported at this moment, given the combined weight of our results and those of B10.

The parameter space of the high-luminosity average GC spectrum in the power-law fit does not overlap with that of the high-luminosity field sources: although the range of $\Gamma$ is consistent, the GC spectrum, while requiring large intrinsic absorption columns, does not appear as absorbed as that of the field sources. In both cases, these large $N_H$ values are consistent with the presence of thermal disk emission; interestingly, the field sources have luminosities larger than those of the GC sources in these two samples, so if they are in a thermal state, they may have smaller residual power-laws in their spectra, which would result in spurious larger $N_H$ (see Brassington et al 2010). The DBB fits give consistent results in the two cases.

The mid-luminosity field and GC power-law parameter spaces are both consistent with $N_H(Gal.)$, therefore dominant intrinsic power-law components are consistent with these results. However, the GC spectrum is harder (flatter power-law), than that of the field sources. The GC spectrum is also consistent with a DBB model and Galactic $N_H$; the field spectrum, instead, is not well fitted by the DBB model, which also would require unphysical low $N_H$, suggesting that this 'average' spectrum include a significant power-law component (B10).

However, since the high and mid luminosity samples typically include only a small number of sources, vagaries of small-object samples compounded with spectral variability may affect our results. Instead, one would expect that these variability effects may be averaged out in the low-luminosity spectra, which in both cases include a large number of sources. We find a suggestive difference in the results of the joint spectral analysis of the large samples of 17 GC and 41 field LMXBs with $L_X < 6 \times 10^{37}$ erg s$^{-1}$. While, as discussed above, the low-luminosity field spectrum is consistent with a power-law dominated emission and Galactic $N_H$, the GC spectrum requires high values of $N_H$. As shown in Fig. 11, the parameter spaces of these sources barely overlap in the power-law fits. The DBB contours of the GC spectrum allow $N_H(Gal)$, which is excluded in the similar fit for the low-luminosity field sources. We discuss possible implications of this result below.

Considering the results of the joined fit of the entire field and GC LMXB samples, we obtain best-fit $\Gamma \sim 1.6 - 1.7$. This range is consistent with previous reports of joint fits of LMXB populations detected with *Chandra* observations in elliptical galaxies (e.g., Irwin et al 2003; Sivakoff et al 2008).

### 6.3. The Elusive Metallicity Effect and differences between GC and Field sources

While it is a well established observational fact that higher metallicity red GCs are more likely to host LMXBs, the effect of metallicity on LMXB formation and evolution is a debated issue (Bellazzini et al 1995; Maccarone et al 2004; Ivanova 2006; see reviews: Fabbiano 2006, Verbunt & Lewin 2006, Kim E. et al 2006). A discriminant among proposed scenarios may be offered by the X-ray spectra of LMXBs in red and blue GCs.

A tendency for LMXBs in metal-rich GCs to have softer spectra was suggested in M31 (Irwin & Bregman 1999) and NGC 4472 (Maccarone et al 2003), but was not verified by the analysis of the X-ray colors of a larger sample of sources culled from six elliptical galaxies observed with *Chandra* (Kim E. et al 2006). Harder spectra in metal-poor GC LMXBs would be expected in the irradiated wind model of Maccarone et al (2004) because of the enhanced absorption associated with these winds, which would also speed up the evolution of the donor star and be responsible for a shorter lifetime of the LMXB (and thus the smaller number of LMXBs associated with blue, metal poor GCs). The different scenario advocated by Ivanova (2006) instead suggests that the relative lack of LMXBs in blue GCs may be related to the lack of outer convective zone in metal-poor main-sequence stars. Without outer convective zones, magnetic braking and thus the formation of an LMXB would not occur. The Ivanova model does not predict a particular spectral signature in the X-ray band. The relative lack of LMXBs in blue GCs could also be due to a smaller stellar encounter rate, resulting from the smaller stellar radii of low metallicity stars (see Berger et al 2006), although the reduction in radius is probably too small to affect strongly the encounter rate; we would not expect a spectral signature in this scenario, either.

Our results show (Fig. 12) that the allowed parameter spaces of the red and blue GC samples spectral fitting (for both power-law and disk models) could be consistent with the presence of larger $N_H$ in the blue sample, as predicted by Maccarone et al (2004). However, this effect is only suggestive in our GC sample. A similar result was found by Sivakoff et al (2008) in NGC 4697.

The differences between the low-luminosity spectral results in field and GC sources discussed in §6.2, if taken at face value, may point to a possible intriguing difference between GC and field sources. If the average metallicity of the GC sources is smaller than that of the field sources, in the Maccarone et al (2004) scenario we may also expect to see larger intrinsic absorption in the GC sample, compared to the field. We note, however, that only five of these 17 GC sources are in blue low-metallicity GCs. Alternatively, these differences may point to physical differences in the nature of accreting binaries formed in the field and in GC, resulting - for the same luminosity range of $L_X < 6 \times 10^{37}$ erg s$^{-1}$ - in accretion-disk dominated systems in GC and power-law dominated systems in the field, as discussed in §6.2. In addition, it is also possible that because of source crowding in GCs, the emission of these fainter GC X-ray sources may be the result of the collective emission of a different population of LMXBs, fainter than that of the field sources, which are likely to be individual LMXBs. This puzzling result will need to be checked with future studies of LMXB populations in other early type galaxies.

### 6.4. LMXB Formation and GC Parameters

As discussed in §5, we find that in NGC 4278 LMXBs tend to be found preferentially in more luminous and red GCs, as observed in other galaxies. These effects, first reported by Angelini et al (2001), Kundu et al (2002), and Sarazin et al (2003) have been confirmed in several studies (see review, Fabbiano 2006), and most recently in the survey of 11 Virgo galaxies by Sivakoff et al (2007).

A surprising result of our analysis of correlations between $L_X$, optical magnitude $V$, and galactocentric distance $R_G$ is the presence of a significant $L_X - R_G$ correlation in the GC-LMXB sample of NGC 4278. This $L_X - R_G$ effect is absent in the field LMXB sample, suggesting that it is an intrinsic property of GC sources in this galaxy. One possible scenario is that these more central GCs have a richer LMXB population than GCs at larger $R_G$. The possibility of multiple sources is allowed by the observed variability pattern, given the statistics of our data. Typically we find that GC sources vary by a factor of a few, so even if a source dominates the emission we cannot exclude the presence of fainter but still luminous LMXBs in a single GC, which cannot be detected individually - even with *Chandra*-, given the distance of NGC 4278 (see light curves in Fig. 3 and in B09); moreover, if the population of LMXBs is richer (effectively, if the XLF is modified by increasing the normalization; see e.g. Kim et al 2009, for a study of LMXB XLF), the larger $L_X$ could be due to a single source, resulting from the increased probability to sample the low-probability, high luminosity tail of the LMXB XLF in these GCs. A second, more difficult to understand, possibility is that there is a change in the slope of the XLF, producing prevalently more luminous sources in GCs at smaller $R_G$.

Our correlation analysis of §5 excludes that the $L_X - R_G$ relation may be a byproduct of a GC luminosity/mass effect (i.e. more luminous/massive GC being found at smaller $R_G$), which would result in a larger fraction of LMXBs formed in richer stellar populations. If it is not due to an underlying luminosity effect, the $L_X - R_G$ correlation could be related to the structure of the GC itself changing with galactocentric radius. If the NGC 4278 GCs tend to be more compact at smaller $R_G$, stellar encounters would be favored, and thus LMXB production. Interestingly, relatively more compact GCs have been found at smaller $R_G$ in galaxies such as the Milky Way, M31, NGC 5128, and the LMC (van den Bergh et al 1991; Hodge 1962; Crampton et al 1985; Hesser et al 1984; see Brodie & Strader 2006). In these galaxies, cluster diameter does not correlate with cluster luminosity, consistent with a lack of luminosity effect, as we have discussed above.

The GC LMXB formation rate depends on the rates of tidal capture and binary exchange which are functions of the size of the GC and the central stellar density ($\Gamma \sim \rho_0^{1.5} r_c^2$; Verbunt & Lewin 2006; Verbunt 2003, where $\rho_0$ is the central stellar density and $r_c$ the core radius; Bregman et al 2006 advocate for the use of the half mass (or light) radius $r_h$ rather than $r_c$ which can be hard to determine). In the Galaxy, LMXBs tend to occur preferentially in more dense and compact GCs (Bregman et al 2006), and a correlation of LMXB occurrence with GC mass and half mass radius is also found in the Virgo Cluster galaxies (Sivakoff et al 2007). Moreover, the observed number of close binaries emitting X-rays in Galactic GCs has been found to be correlated with $\Gamma$, and the number of LMXBs is also expected to be correlated with $\Gamma$ (Pooley et al 2003). For a given GC luminosity (there is no observed $V$ dependence), we would have $\Gamma \sim r_c^{-2.5}$ and we could then approximate the functional dependence of these dynamical effects as $\Gamma \sim R_G^{-1.2}$, using the square-root dependence between GC diameter and $R_G$ of van den Berg et al (1991). A function of this type could be consistent with the $L_X - R_G$ relation observed in NGC 4278, considering the range of approximations made in deriving the $\Gamma - R_G$ relation and the wide observed scatter of fig. 13. If this scenario holds, the $L_X - R_G$ relation may be due to a larger number of LMXBs in GCs at smaller $R_G$ (increase in the XLF normalization), rather that to intrinsically luminous single LMXBs being prevalently formed in GCs closer to the central regions of the galaxy (flattening of the XLF slope).

## 7. Summary and Conclusions

We have presented the results of a detailed spectral analysis of the population of LMXBs detected in the elliptical galaxy NGC 4278 with a series of six *Chandra* ACIS-S observations. These sources are extracted from the catalog of B09.

1) Seven of these sources, all with luminosity in B09 at or above the Eddington limit for an accreting NS, were studied individually. Four of these luminous sources are associated with GCs in NGC 4278; the other three do not have optical counterparts and are found in the stellar field of NGC 4278 (see B09). The statistics of our data only allow meaningful fits to single emission models, either power-law or thermal-disk (DBB), with absorption columns. The DBB fits result in best-fit disk temperatures in the range 0.7-1.7 keV, consistent with those measured in BHBs (see McClintock & Remillard 2006); the power-law fits give a range of $\Gamma \sim 1.2-2.0$, overlapping those observed in BHBs, although somewhat flatter. Steeper power-laws tend to be associated with higher $N_H$, exceeding the line of sight Galactic value. This result could be symptomatic of the presence of thermal disk emission (B10), suggesting that in this instance the binary is in a thermal disk-dominated state. Some of these LMXBs undergo spectral/luminosity variability, reminiscent of those observed in BHBs (see Remillard & McClintock 2006). Using the best-fit spectral parameters, we find a range of luminosities for these sources $\sim 3-8\times10^{38}$erg s$^{-1}$, suggesting that they may all be BHBs. This applies also to the GC-LMXBs, increasing significantly the number of BHB candidates in GCs.The BH masses derived from a comparison of our spectral results with the $L_X \sim T_{in}^4$ of Galactic BHBs are in the 5-15 $M_\odot$ range, as observed in the Milky Way.

2) The analysis of joint spectra of sources selected in three luminosity ranges ($L_X \geq 1.5\times10^{38}$ erg s$^{-1}$, which includes the sources also analyzed individually; $L_X$ in the range $6\times10^{37}-1.5\times10^{38}$ erg s$^{-1}$; and $L_X < 6\times10^{37}$ erg s$^{-1}$), suggests that the high luminosity sources are likely to have a prominent thermal disk emission component in their spectra, while power-law components are likely to be important in the mid and low-luminosity spectra.

3) We find a marginal, but not statistically significant, difference in the average spectral properties of sources in red and blue GCs, in the sense of red GC LMXBs having spectral fit parameter spaces allowing for lower $N_H$ values. Although this $N_H$ effect is in the sense of the predictions of the wind-irradiated model of metal-poor blue GC stars (Maccarone et al 2004), given the significance of our results we cannot add to the discussion either in support or against this model. Future analyses of even larger LMXB samples may help resolve this issue.

4) We find, however, a possible difference in the spectral results of the low-luminosity field and GC samples, suggesting higher $N_H$, in excess of the Galactic line of sight column, in the latter, while the former could be fitted with $N_H(Gal.)$. We speculate that this may be a metallicity effect, as in the Maccarone at al (2004) model.Alternatively, these differences may point to intrinsic physical differences between field (power-law dominated) and GC LMXBs (accretion-disk dominated). Future studies with larger LMXB samples are needed to confirm the importance and the generality of this result.

5) Analysis of average sample properties and correlations confirm many effects reported in studies of other galaxies (see Fabbiano 2006 and refs. therein), including the prevalent association of X-ray sources with more luminous and red GCs.

6) We report a significant $L_X - R_G$ correlation (where $R_G$ is the galactocentric radius) in the GC-LMXB sample of NGC 4278. This correlation is absent in the field LMXB sample, suggesting that it reflects an intrinsic property of the GC population. We can exclude that the $L_X - R_G$ correlation is the by-product of a luminosity effect, and suggest that it may be related to the presence of more compact GCs at smaller galactocentric radii, fostering more effective binary formation.The increased $L_X$ in more central GCs may be due to richer LMXB populations in these GCs. Although these luminous GC LMXBs vary, the observed patterns of source variability do not exclude the presence of multiple sources in a given cluster. Moreover, enhanced LMXB formation would increase the probability of detecting rarer high luminosity sources.

This work was supported by the *Chandra* GO grant G06-7079A (PI: Fabbiano) and NASA contract NAS8-39073 (CXC). This paper is partly based on work performed by L. Lentati while visiting CfA as part of a visiting student program sponsored by Southampton University, and these results were included in his Master Thesis presentation. A. Richings also acknowledges support by the Southampton University CfA visiting program. The data analysis was supported by the CXC CIAO software and CALDB. We have used the NASA NED and ADS facilities, and have extracted archival data from the *Chandra* archives.

*Facilities:* CXO (ACIS).

## Appendix A - Spectral Fits of Individual Sources Outside the HST Field of View

Six additional sources within the $D_{25}$ ellipse of NGC 4278 were detected with $> 400$ counts in the coadded data. These sources lie in a region not covered by *HST* observations (see B09), so we do not have information regarding optical counterparts. We report the results of the spectral analysis of these sources in Table 11 and Figs. 17 and 18. The same approach was used as described in § 3. We refer to B09 for information about these sources, including light-curves. Two of these sources (S13 and S228) vary significantly (see light curves in Fig. 16), however there are too few counts in the Groups 1 spectra of both sources to constrain the spectral parameters usefully, therefore in Table 11 we report only the results obtained by freezing $N_H$ to the Galactic line of sight value. For S13 we find that leaving both parameters free to vary, the confidence contours are consistent with cooler DBB emission or steeper and intrinsically absorbed PO (Fig. 17); for S228, the contours suggest significantly higher $N_H$, as well as cooler emission in the DBB fit (see Fig. 18). Comparison with the B10 simulation suggests that of the sources that are not found to vary, S31, S41, and S42 may be in thermal state, in that the PO fits suggest absorbed steep power-laws; S5 is likely to have a significant power-law component.

---

Table 1. Observation Log

| Obs No. | OBSID | Date | $T_{eff}$ (Ks) |
|---|---|---|---|
| 1 | 4741 | 2005-02-03 | 37 |
| 2 | 7077 | 2006-03-16 | 108 |
| 3 | 7078 | 2006-07-25 | 48 |
| 4 | 7079 | 2006-10-24 | 103 |
| 5 | 7081 | 2007-02-20 | 107 |
| 6 | 7080 | 2007-04-20 | 55 |

Table 2. Properties of Single component fits from Two component simulations

| $kT_2$ or $\Gamma_2$ | Single PO fit | | | | Single DBB fit | | | |
|---|---|---|---|---|---|---|---|---|
| | 40% | 50% | 75% | 90% | 40% | 50% | 75% | 90% |
| | $N_H/\Gamma_2$ | $N_H/\Gamma_2$ | $N_H/\Gamma_2$ | $N_H/\Gamma_2$ | $N_H/kT_2$ | $N_H/kT_2$ | $N_H/kT_2$ | $N_H/kT_2$ |
| PO input model, $\Gamma_1$=1.7 | | | | | | | | |
| 1.2 | 1.4 / 1.5 | 1.5 / 1.5 | 1.3 / 1.3 | 1.8 / 1.3 | 0.0 / 1.4 | 0.0 / 1.5 | 0.0 / 1.9 | 0.0 / 2.1 |
| 1.5 | 2.0 / 1.6 | 2.1 / 1.6 | 1.9 / 1.6 | 2.1 / 1.5 | 0.0 / 1.1 | 0.0 / 1.2 | 0.0 / 1.3 | 0.0 / 1.3 |
| 1.7 | 2.1 / 1.7 | 2.0 / 1.7 | 2.2 / 1.7 | 1.7 / 1.7 | 0.0 / 1.0 | 0.0 / 1.1 | 0.0 / 1.0 | 0.0 / 1.1 |
| 2.1 | 1.7 / 1.9 | 1.6 / 1.9 | 2.2 / 2.0 | 2.1 / 2.1 | 0.0 / 0.9 | 0.0 / 0.9 | 0.0 / 0.7 | 0.0 / 0.7 |
| 2.3 | 1.2 / 1.9 | 1.4 / 2.0 | 1.6 / 2.2 | 2.0 / 2.3 | 0.0 / 0.8 | 0.0 / 0.8 | 0.0 / 0.7 | 0.0 / 0.5 |
| DBB input model, $kT_1$=1.0 keV | | | | | | | | |
| 0.50 | 19.8 / 2.6 | 20.0 / 2.7 | 23.3 / 3.1 | 25.4 / 3.3 | 0.6 / 0.7 | 0.4 / 0.7 | 1.0 / 0.6 | 1.5 / 0.5 |
| 1.00 | 19.1 / 2.2 | 18.0 / 2.1 | 17.9 / 2.1 | 18.7 / 2.2 | 1.9 / 1.0 | 1.7 / 1.0 | 1.7 / 1.0 | 1.9 / 1.0 |
| 1.50 | 17.0 / 2.0 | 16.4 / 1.9 | 15.1 / 1.8 | 13.7 / 1.7 | 2.0 / 1.1 | 1.7 / 1.2 | 1.8 / 1.3 | 1.5 / 1.4 |
| 2.25 | 17.7 / 1.8 | 13.7 / 1.7 | 10.6 / 1.5 | 9.6 / 1.3 | 1.2 / 1.3 | 1.1 / 1.4 | 1.0 / 1.8 | 1.3 / 2.0 |

Note. — Percentages are the flux ratios of the second spectral component to the total. The absorption column $N_H$ is given in units of $10^{20}$cm$^{-2}$. For the two component PO model the output $N_{H,\Gamma}$ values have a variance of $\sim$2.0 and $\Gamma$ variance of 0.10. The $N_{H,\mathrm{DBB}}$ variance is $\sim$0.0 with a kT variance of 0.13. For the two component DBB models, the output $N_{H,\Gamma}$ values have a variance of $\sim$3.0 and $\Gamma$ variance of 0.15. The $N_{H,\mathrm{DBB}}$ variance is $\sim$1.2 with a kT variance of 0.8.

Table 3. Single GC Fit Results

| Source | Obs | Counts | Count Rate ($10^{-3}$ $s^{-1}$) | Model | Statistic C-stat | $\chi^2/\nu$ | $\nu$ | $P_{Cash}$ (%) | $P_{\chi^2}$ (%) | $N_H$ ($10^{20} cm^{-2}$) | $\Gamma$ | kT (keV) | $L_X$ ($10^{38}$ erg$s^{-1}$) |
|---|---|---|---|---|---|---|---|---|---|---|---|---|---|
| 96 | All | 826.4±28.7 | 1.3±0.3 | PO | - | 1.4 | 25 | - | 0.08 | 8.8±7.3 | 1.5±0.2 | - | 5.05 |
| | | | | DBB | - | 1.5 | 26 | - | 0.05 | F | - | 1.4±0.3 | 4.11 |
| 163 | Group 1 | 124.2±11.1 | 1.5±0.4 | PO | 370 | - | - | 89 | - | 27.0±13.8 | 2.5±0.5 | - | 5.48 |
| | (1+3) | | | DBB | 368 | - | - | 93 | - | 8.8±6.0 | - | 0.7±0.2 | 2.61 |
| | Group 2 | 453.7±21.3 | 1.2±0.1 | PO | 1042 | - | - | 57 | - | 11.0±4.3 | 1.4±0.1 | - | 4.09 |
| | (2+4+5+6) | | | DBB | 1036 | - | - | 65 | - | F | - | 1.6±0.2 | 3.16 |
| | Group 3 | 267.0±16.3 | 1.3±0.1 | PO | 572 | - | - | 60 | - | 9.3±5.2 | 1.4±0.2 | - | 4.17 |
| | (2+4) | | | DBB | 569 | - | - | 70 | - | F | - | 1.6±0.2 | 3.29 |
| | Group 4 | 186.7±13.7 | 1.2±0.1 | PO | 470 | - | - | 52 | - | 14.4±7.5 | 1.5±0.3 | - | 4.06 |
| | (5+6) | | | DBB | 467 | - | - | 71 | - | 3.9±3.7 | - | 1.6±0.4 | 3.14 |
| | All Obs | 577.9±21 | 1.3±0.2 | PO | 1419 | - | - | 55 | - | 12.9±3.8 | 1.6±0.1 | - | 4.14 |
| | | | | DBB | 1411 | - | - | 68 | - | 2.6±2.6 | - | 1.4±0.2 | 3.04 |
| 185 | All Obs | 550.4±24.0 | 1.2±0.1 | PO | 1423 | - | - | 44 | - | 12.5±4.3 | 1.7±0.2 | - | 3.50 |
| | | | | DBB | 1357 | - | - | 74 | - | 2.7±2.7 | - | 1.2±0.1 | 2.53 |
| 194 | Group 1 | 150.3±11 | 1.8±0.2 | PO | 431 | - | - | 78 | - | 23.3±10.0 | 1.8±0.4 | - | 6.00 |
| | (1+3) | | | DBB | 427 | - | - | 89 | - | 9.7±5.9 | - | 1.1±0.1 | 3.94 |
| | Group 2 | 121.9±11 | 1.1±0.1 | PO | 250 | - | - | 45 | - | 3.9±3.9 | 1.9±0.3 | - | 2.81 |
| | (2) | | | DBB | 251 | - | - | 98 | - | F | - | 0.7±0.1 | 1.87 |
| | Group 3 | 307.4±16 | 1.1±0.1 | PO | 702 | - | - | 49 | - | 5.7±4.1 | 1.6±0.2 | - | 3.25 |
| | (4+5+6) | | | DBB | 694 | - | - | 90 | - | F | - | 1.1±0.1 | 2.29 |
| | All Obs | 579.6±24 | 1.3±0.4 | PO | - | 0.65 | 14 | - | 0.82 | F | 1.3±0.1 | - | 3.60 |
| | | | | DBB | - | 0.74 | 15 | - | 0.75 | F | - | 1.5±0.2 | 3.09 |

Note. — Based on the B10 simulations (see text) a significant/strong DBB component is expected in Group 1 spectra of S163 and S194, and in S185.

Table 4. Single Field Fit Results

| Source | Obs | Counts | Count Rate ($10^{-3}\ s^{-1}$) | Model | Statistic C-stat | $\chi^2/\nu$ | $\nu$ | $P_{Cash}$ (%) | $P_{\chi^2}$ (%) | $N_H$ ($10^{20} cm^{-2}$) | $\Gamma$ | kT (keV) | $L_X$ ($10^{38}$ erg$s^{-1}$) |
|---|---|---|---|---|---|---|---|---|---|---|---|---|---|
| 146 | Group 1 | 861.7±29.4 | 2.05±0.5 | PO | - | 0.61 | 29 | - | 0.95 | 12.6±6.1 | 1.4±0.2 | - | 8.19 |
| | (1+3+4+5+6) | | | DBB | - | 0.58 | 30 | - | 0.97 | F | - | 1.6±0.2 | 6.15 |
| | Group 2 | 154.3±12.4 | 1.4±0.1 | PO | 247 | - | - | 33 | - | 19.9±7.5 | 2.1±0.3 | - | 4.77 |
| | (2) | | | DBB | 246 | - | - | 71 | - | 3.9±3.9 | - | 1.0±0.2 | 2.90 |
| | All Obs | 1016.2±31.9 | 2.22±0.6 | PO | - | 0.67 | 35 | - | 0.93 | 15.3±5.7 | 1.6±0.2 | - | 6.97 |
| | | | | DBB | - | 0.60 | 36 | - | 0.97 | F | - | 1.5±0.1 | 5.23 |
| 158 | Group 1 | 463.7±21.5 | 2.47±0.3 | PO | 801 | - | - | 58 | - | 20.7±5.1 | 1.9±0.2 | - | 8.56 |
| | (1+3+4) | | | DBB | 790 | - | - | 68 | - | 5.3±3.7 | - | 1.2±0.1 | 5.66 |
| | Group 2 | 244.8±15.6 | 2.27±0.2 | PO | 299 | - | - | 58 | - | 22.7±6.2 | 2.2±0.2 | - | 7.99 |
| | (2) | | | DBB | 293 | - | - | 82 | - | 5.9±3.8 | - | 0.9±0.1 | 4.55 |
| | Group 3 | 443.2±21.1 | 2.72±0.3 | PO | - | 0.95 | 15 | - | 0.50 | 8.9±4.0 | 1.7±0.1 | - | 7.72 |
| | (5+6) | | | DBB | - | 1.05 | 16 | - | 0.39 | F | - | 1.2±0.1 | 6.98 |
| | All Obs | 1151.7±33.9 | 2.51±0.7 | PO | - | 1.23 | 41 | - | 0.14 | 21.4±4.0 | 1.9±0.2 | - | 7.00 |
| | | | | DBB | - | 1.25 | 41 | - | 0.13 | 1.9±1.9 | - | 1.2±0.1 | 4.91 |
| 184 | Group 1 | 241.0±15.5 | 2.60±0.5 | PO | 532 | - | - | 53 | - | 13.1±5.6 | 1.6±0.2 | - | 8.72 |
| | (1+6) | | | DBB | 533 | - | - | 72 | - | 2.2±2.2 | - | 1.6±0.2 | 6.72 |
| | Group 2 | 557.8±23.6 | 2.2±0.4 | PO | - | 1.11 | 19 | - | 0.33 | 4.5±4.5 | 1.3±0.2 | - | 6.72 |
| | (2+3+4) | | | DBB | - | 1.10 | 20 | - | 0.34 | F | - | 1.7±0.3 | 5.72 |
| | Group 3 | 290.0±17.0 | 2.70±0.2 | PO | 366 | - | - | 83 | - | 25.5±7.0 | 1.8±0.2 | - | 10.36 |
| | (5) | | | DBB | 374 | - | - | 94 | - | 8.0 ±4.2 | - | 1.4±0.2 | 7.16 |
| | All Obs | 1090.4±33.0 | 2.4±0.6 | PO | - | 1.25 | 38 | - | 0.14 | 5.0±2.7 | 1.4±0.1 | - | 6.96 |
| | | | | DBB | - | 1.35 | 39 | - | 0.07 | F | - | 1.5±0.2 | 5.47 |

Note. — Based on the B10 simulations (see text) a significant/strong DBB component is expected in S146, S158 and Group 1 and Group 3 spectra of S184.

Table 5. GC Luminosity Samples

| Sample | Counts | Model | $\nu$ | $\chi^2/\nu$ | $P_{\chi^2}$ (%) | $N_H$ ($10^{20} cm^{-2}$) | $\Gamma$ | kT (keV) |
|---|---|---|---|---|---|---|---|---|
| High | 2027.9±45.0 | PO | 84 | 1.3 | 0.02 | $7.7^{+2.3}_{-1.7}$ | 1.5±0.1 | - |
| L $\geq 1.5\text{x}10^{38}\text{erg}s^{-1}$ | | PO | 85 | 1.4 | 0.009 | F | 1.3±0.1 | - |
| | | DBB | 85 | 1.4 | 0.008 | F | - | 1.4±0.1 |
| Mid | 1372.4 ±37.0 | PO | 60 | 1.02 | 0.42 | $8.0^{+12.3}_{-8.0}$ | 1.5±0.1 | - |
| $6\text{x}10^{37} \geq$ L $< 1.5\text{x}10^{38}$ erg$s^{-1}$ | | PO | 61 | 0.95 | 0.59 | F | 1.4±0.1 | - |
| | | DBB | 61 | 1.03 | 0.41 | F | - | 1.4±0.1 |
| High + Mid | 3400.3±55.9 | PO | 155 | 1.2 | 0.03 | 7.6±1.1 | 1.50±0.04 | - |
| L $\geq 6\text{x}10^{37}$ erg$s^{-1}$ | | DBB | 156 | 1.4 | 0.002 | F | - | 1.5±0.1 |
| Low | 1030.0±32.1 | PO | 64 | 0.99 | 0.49 | 16.7±4.9 | $2.0^{+0.2}_{-0.1}$ | - |
| L $< 6\text{x}10^{37}$ erg$s^{-1}$ | | DBB | 65 | 1.07 | 0.34 | F | - | $1.08^{+0.08}_{-0.09}$ |
| All | 4430.3±69 | PO | 212 | 1.1 | 0.07 | 10.3±1.6 | 1.6±0.1 | - |
| | | DBB | 213 | 1.2 | 0.02 | F | - | 1.3±0.1 |

Table 6. Field Luminosity Samples

| Sample | Counts | Model | $\nu$ | $\chi^2/\nu$ | $P_{\chi^2}$ (%) | $N_H$ ($10^{20} cm^{-2}$) | $\Gamma$ | kT (keV) |
|---|---|---|---|---|---|---|---|---|
| High | 3353.8±57.9 | PO | 142 | 1.05 | 0.33 | $14.3^{+13.1}_{-15.0}$ | $1.70^{+0.04}_{-0.03}$ | - |
| L ≥ 1.5x10$^{38}$ erg$s^{-1}$ | | DBB | 143 | 1.02 | 0.41 | F | - | $1.38^{+0.04}_{-0.07}$ |
| Mid | 1641.9±40.5 | PO | 69 | 0.85 | 0.82 | 5.4±2.1 | 1.82±0.06 | - |
| 6x10$^{37}$ ≥ L < 1.5x10$^{38}$ erg$s^{-1}$ | | DBB | 70 | 1.56 | 0.002 | F | - | $0.96^{+0.05}_{-0.06}$ |
| Low | 2027.2±45.0 | PO | 132 | 0.92 | 0.73 | 4.2±2.3 | 1.6±0.1 | - |
| L < 6x10$^{37}$ erg$s^{-1}$ | | DBB | 133 | 1.1 | 0.14 | F | - | 1.2±0.1 |
| All[a] | 8851.2±94.1 | PO | 406 | 1.09 | 0.09 | 8.9±1.4 | $1.72^{+0.02}_{-0.05}$ | - |
| | | DBB | 407 | 1.29 | 0.00001 | F | - | $1.23^{+0.02}_{-0.03}$ |

Note. — [a] The number of counts is larger than the sum of the counts of the three subsamples, because 'All' contains additional sources, see §2

Table 7. GC Color Samples

| Sample | Counts | Model | $\nu$ | $\chi^2/\nu$ | $P_{\chi^2}$ (%) | $N_H$ ($10^{20} cm^{-2}$) | $\Gamma$ | kT (keV) |
|---|---|---|---|---|---|---|---|---|
| Red | 3616.7±60.1 | PO | 179 | 0.95 | 0.67 | 8.2±1.4 | 1.6±0.1 | - |
| V-I ≥ 1.05 | | DBB | 180 | 1.2 | 0.07 | F | - | 1.4±0.1 |
| Blue | 1290.3±35.9 | PO | 58 | 0.80 | 0.86 | 14.7±4.4 | 1.7±0.1 | - |
| V-I < 1.05 | | DBB | 59 | 0.85 | 0.79 | F | - | 1.4±0.1 |

Table 8. Fraction of GCs associated with an LMXB, as a function of $V_{mag}$

| $V_{mag}$ | No. (GC) | | No. (GC-LMXB) | | f (%) | |
|---|---|---|---|---|---|---|
| | Red | Blue | Red | Blue | Red | Blue |
| > 22.5 | 90 | 106 | 9 | 6 | 10±3 | 6±2 |
| ≤ 22.5 | 31 | 39 | 16 | 8 | 51±16 | 21±8 |

Table 9. GC Color and $L_X$ statistics.

| Color | $L_X$ | Sources | f (%) |
|---|---|---|---|
| Red | High | 3 | 12±7 (14) |
| | Mid | 8 (7) | 32±13 (29) |
| | Low | 14 (13) | 56 ±19 (57) |
| | Total | 25 (23) | 100 |
| Blue | High | 2 | 14±4 (17) |
| | Mid | 4 (3) | 29±16 (25) |
| | Low | 8 (7) | 57 ±25 (58) |
| | Total | 14 (12) | 100 |

Note. — Values in parentheses are the number of sources, and related fraction, with flux determined at $\geq 3\sigma$ confidence.

Table 10. Spearmans Rank Probabilities

| Correlation | N | SR coefficient | Probability |
|---|---|---|---|
| $L_X$ - R (GC all) | 39 | -0.49 | 0.0008 |
| $L_X$ - R (GC $3\sigma$) | 35 | -0.51 | 0.004 |
| $L_X$ - R (Field all) | 58 | -0.22 | 0.05 |
| $L_X$ - R (Field $3\sigma$) | 43 | -0.19 | 0.11 |
| $V_{mag}$ - R (GC all) | 39 | -0.10 | 0.27 |
| $V_{mag}$ - R (GC $3\sigma$) | 35 | -0.15 | 0.20 |
| $L_X$ - $V_{mag}$ (GC all) | 39 | -0.020 | 0.11 |
| $L_X$ - $V_{mag}$ (GC $3\sigma$) | 35 | -0.11 | 0.27 |

Table 11: Best Fit Parameters for Sources Not in HST F.o.V

| Source | Obs | Counts | Count Rate ($10^{-3} s^{-1}$) | Model | Statistic C-stat | $\chi^2/\nu$ | $\nu$ | $P_{cash}$ (%) | $P_{\chi^2}$ (%) | $N_H$ ($10^{20} cm^{-2}$) | $\Gamma$ | $kT$ (keV) | $L_X$ ($10^{38} ergs\ s^{-1}$) |
|---|---|---|---|---|---|---|---|---|---|---|---|---|---|
| 5 | All | 723.6 ± 28.2 | 1.6 ± 0.1 | PO | - | 0.78 | 22 | - | 75 | 1.8F | 1.65 ± 0.12 | - | 3.75 |
| | | | | DBB | - | 1.32 | 22 | - | 15 | 1.8F | - | 0.96 ± 0.13 | 2.67 |
| 13 | Group 1 | 49.5 ± 7.4 | 1.3 ± 0.2 | PO | 175 | - | - | 76 | - | 1.8F | 1.89±0.40 | - | 2.69 |
| | (1) | | | DBB | 169 | - | - | 82 | - | 1.8F | - | 0.56 ± 0.15 | 1.79 |
| | Group 2 | 533.2 ± 23.6 | 3.4 ± 0.2 | PO | - | 1.04 | 19 | - | 40.3 | 3.2 ± 3.2 | 1.67 ± 0.17 | - | 8.59 |
| | (2+3) | | | DBB | - | 1.49 | 20 | - | 7.4 | 1.8F | - | 0.95 ± 0.10 | 5.95 |
| | Group 3 | 521.4 ± 23.7 | 2.0 ± 0.1 | PO | - | 0.40 | 18 | - | 98 | 1.8F | 1.54 ± 0.11 | - | 5.22 |
| | (4+5+6) | | | DBB | - | 1.17 | 18 | - | 28 | 1.8F | - | 1.16 ± 0.15 | 3.85 |
| | All | 1104.1 ± 34.2 | 2.4 ± 0.1 | PO | - | 0.74 | 39 | - | 88.4 | 1.8F | 1.59 ± 0.07 | - | 6.15 |
| | | | | DBB | - | 1.34 | 39 | - | 7.8 | 1.8F | - | 1.04 ± 0.09 | 4.34 |
| 31 | All | 454.0 ± 22.2 | 1.0 ± 0.1 | PO | 1235 | - | - | 52 | - | 25.6 ± 6.3 | 2.54 ± 0.28 | - | 3.51 |
| | | | | DBB | 1225 | - | - | 100 | - | 6.0 ± 4.8 | - | 0.74 ± 0.10 | 1.60 |
| 41 | All | 660.4 ± 26.5 | 1.4 ± 0.1 | PO | - | 0.61 | 16 | - | 88 | 21.3 ± 12.1 | 1.64 ± 0.29 | - | 4.21 |
| | | | | DBB | - | 0.63 | 16 | - | 86 | 3.3 ± 3.3 | - | 1.44 ± 0.32 | 3.14 |
| 42 | All | 429.7 ± 21.6 | 0.9 ± 0.1 | PO | 1145 | - | - | 26 | - | 19 ± 6.4 | 2.28 ± 0.28 | - | 2.80 |
| | | | | DBB | 1127 | - | - | 99 | - | 2.0 ± 2.0 | - | 0.81 ± 0.10 | 1.48 |
| 228 | Group 1 | 88.6 ± 10.8 | 0.30 ± 0.04 | PO | 408 | - | - | 23 | - | 1.8F | 1.70F | - | 0.70 |
| | (1+2+3+4) | | | DBB | 405 | - | - | 62 | - | 1.8F | - | 1.00F | 0.58 |
| | Group 2 | 126.7 ± 11.5 | 2.3 ± 0.2 | PO | 281 | - | - | 95 | - | 18.8 ± 12.4 | 1.95 ± 0.45 | - | 7.13 |
| | (5) | | | DBB | 277 | - | - | 99 | - | 1.8F | - | 1.06 ± 0.22 | 4.30 |
| | Group 3 | 275.0 ± 16.9 | 2.6 ± 0.2 | PO | - | 1.45 | 9 | - | 16.2 | 20.3 ± 14.0 | 1.55 ± 0.14 | - | 8.11 |
| | (6) | | | DBB | - | 1.50 | 9 | - | 14.2 | 4.7 ± 4.7 | - | 1.53 ± 0.37 | 6.13 |
| | All | 490.3 ± 23.1 | 1.07 ± 0.1 | PO | 1060 | - | - | 57 | - | 18.0 ± 5.7 | 1.63 ± 0.17 | - | 5.59 |
| | | | | DBB | 1062 | - | - | 78 | - | 5.9 ± 3.4 | - | 1.44 ± 0.22 | 4.08 |

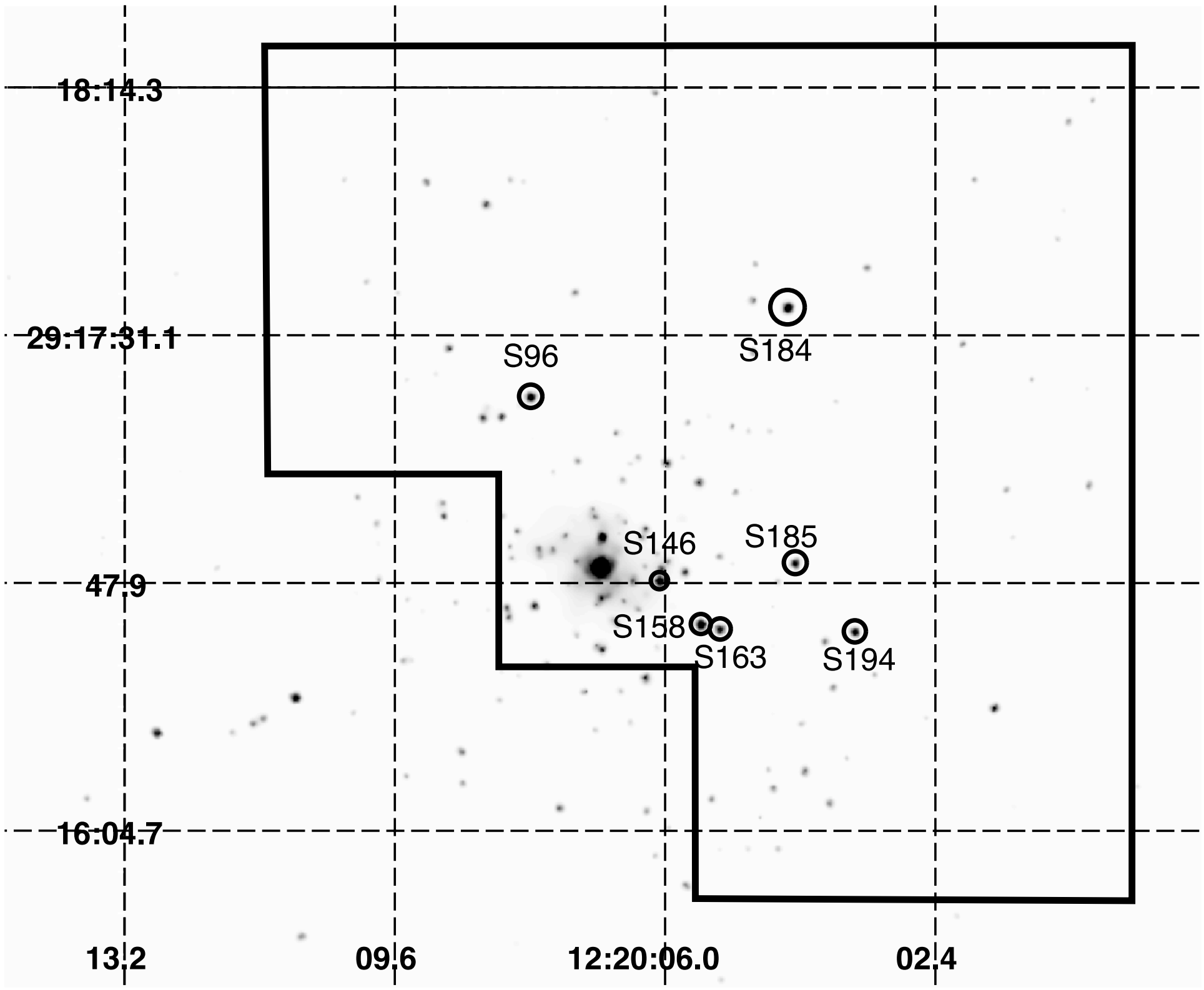

Fig. 1.— *Chandra* image of NGC-4278 with the sources selected for individual analysis and the outline of the *HST* field of view. S96, S163, S185 and S194 are associated with GCs.

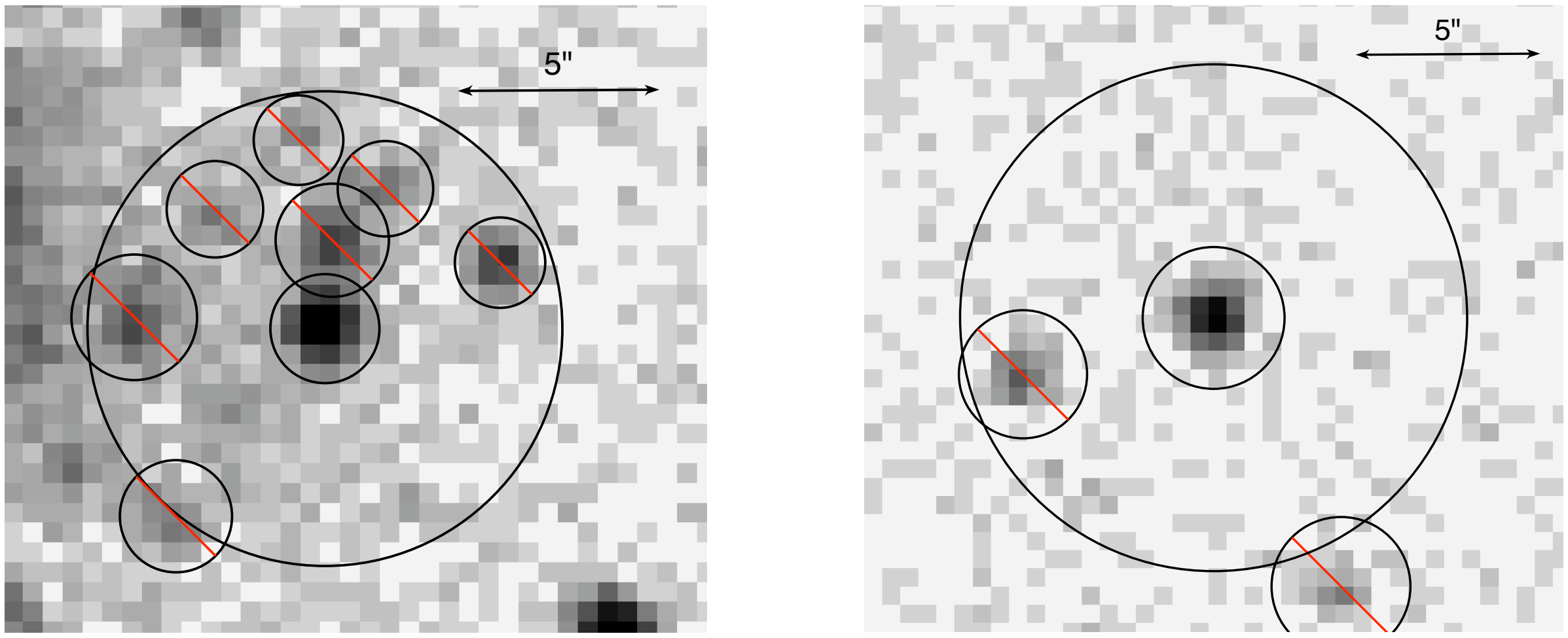

Fig. 2.— Left: source and background extraction regions for S146 (left). Right: same for S194. Regions excluded because of extraneous source contamination are indicated with crossed circles.

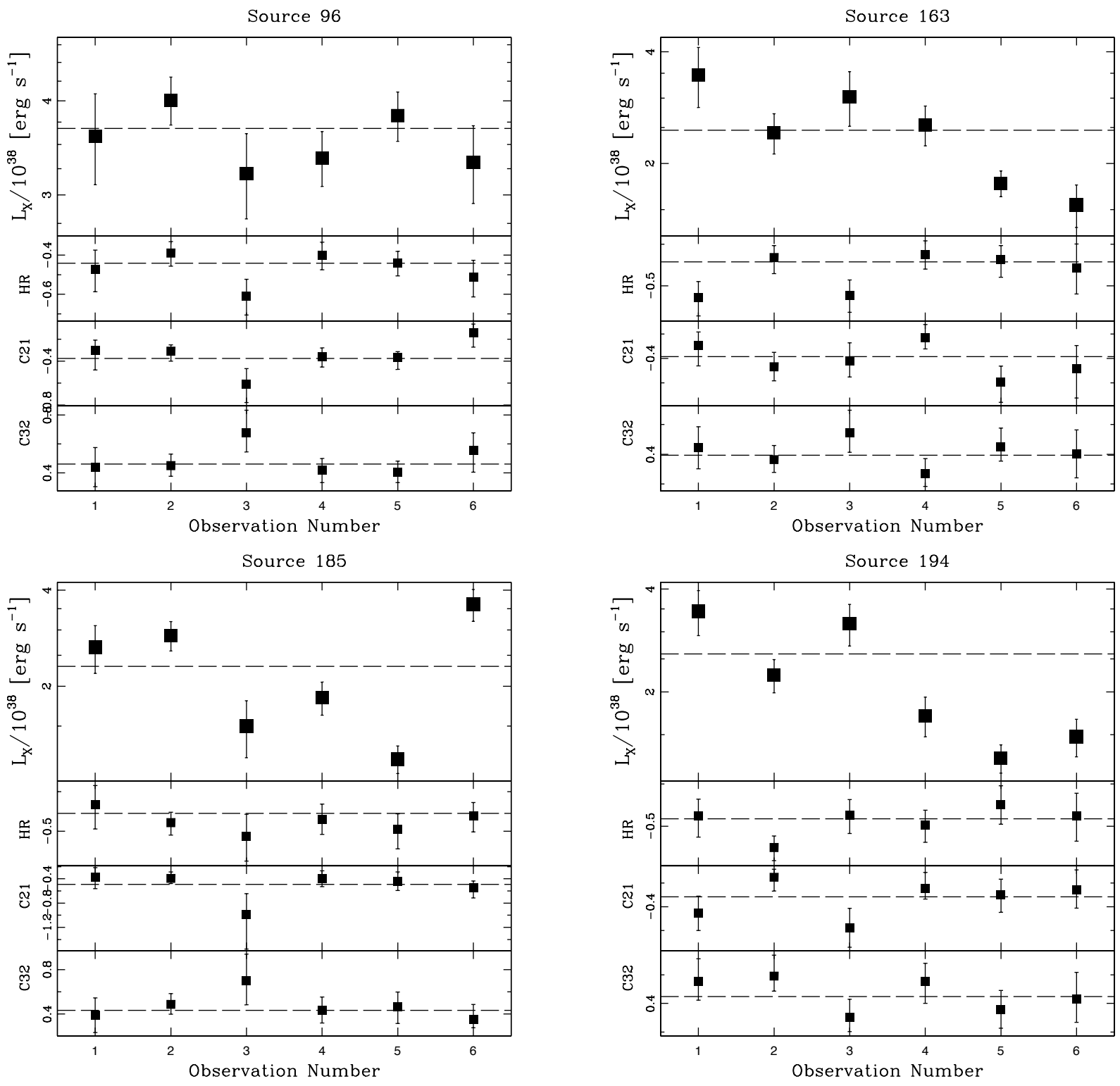

Fig. 3.— Light curves of S96, S163, S185 and S194 from B09, indicating luminosity and spectral variability. Each point represents a separate observation. From the top down: (0.3-8.0) keV $L_X$; hardness ratio, defined as $HR = H - S/(H+S)$ where $H$ is the number of counts in the hard band (2.0-8.0) keV and S is the number of counts in the soft band (0.5-2.0) keV; the color $C21 = logS2 + logS1$; and the color $C32 = logS2 - logH$, where $S21 = 0.3 - 0.9$ keV, $S2 = 0.9 - 2.5$ keV and $H = 2.5 - 8.0 keV$ (see B09)

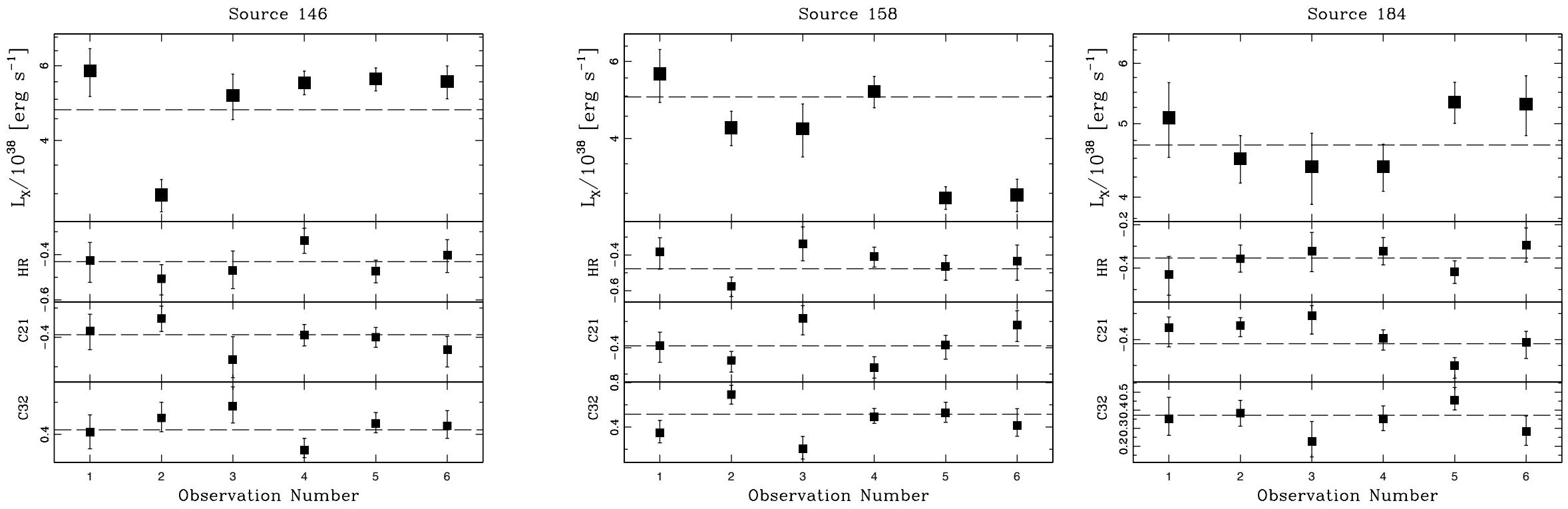

Fig. 4.— Light curves of S146, S158, and S184 from B09, indicating luminosity and spectral variability. Each point represents a separate observation. From the top down: (0.3-8.0) keV $L_X$; hardness ratio, defined as $HR = H - S/(H+S)$ where $H$ is the number of counts in the hard band (2.0-8.0) keV and S is the number of counts in the soft band (0.5-2.0) keV; the color $C21 = logS2 + logS1$; and the color $C32 = logS2 - logH$, where $S21 = 0.3 - 0.9$ keV, $S2 = 0.9 - 2.5$ keV and $H = 2.5 - 8.0keV$ (see B09)

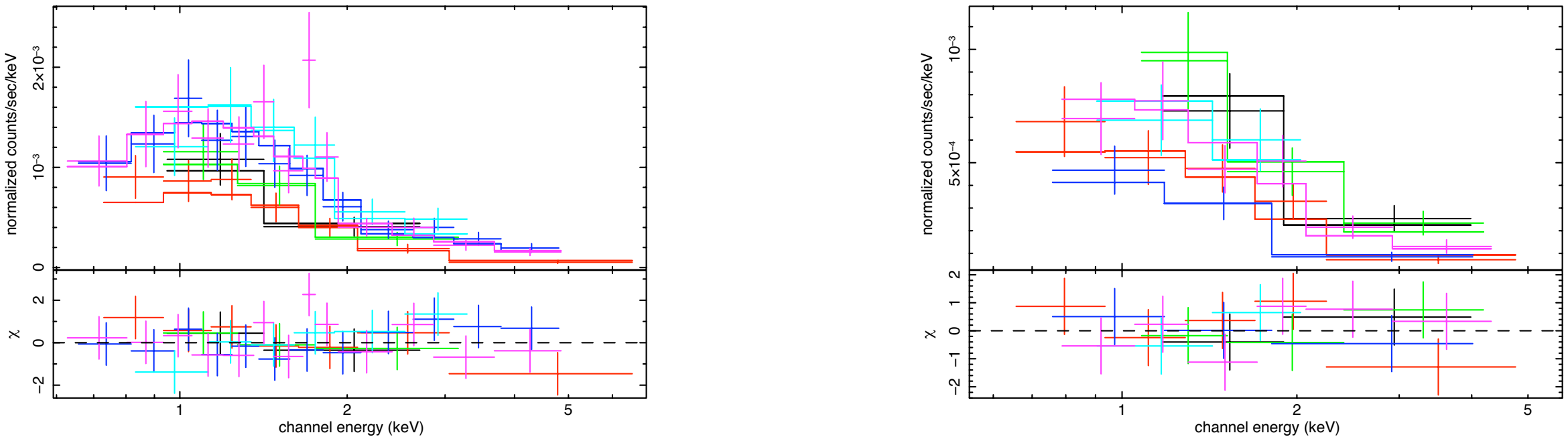

Fig. 5.— Spectra, best-fit model and residuals for the power law fits S146 (left) and S194 (right)

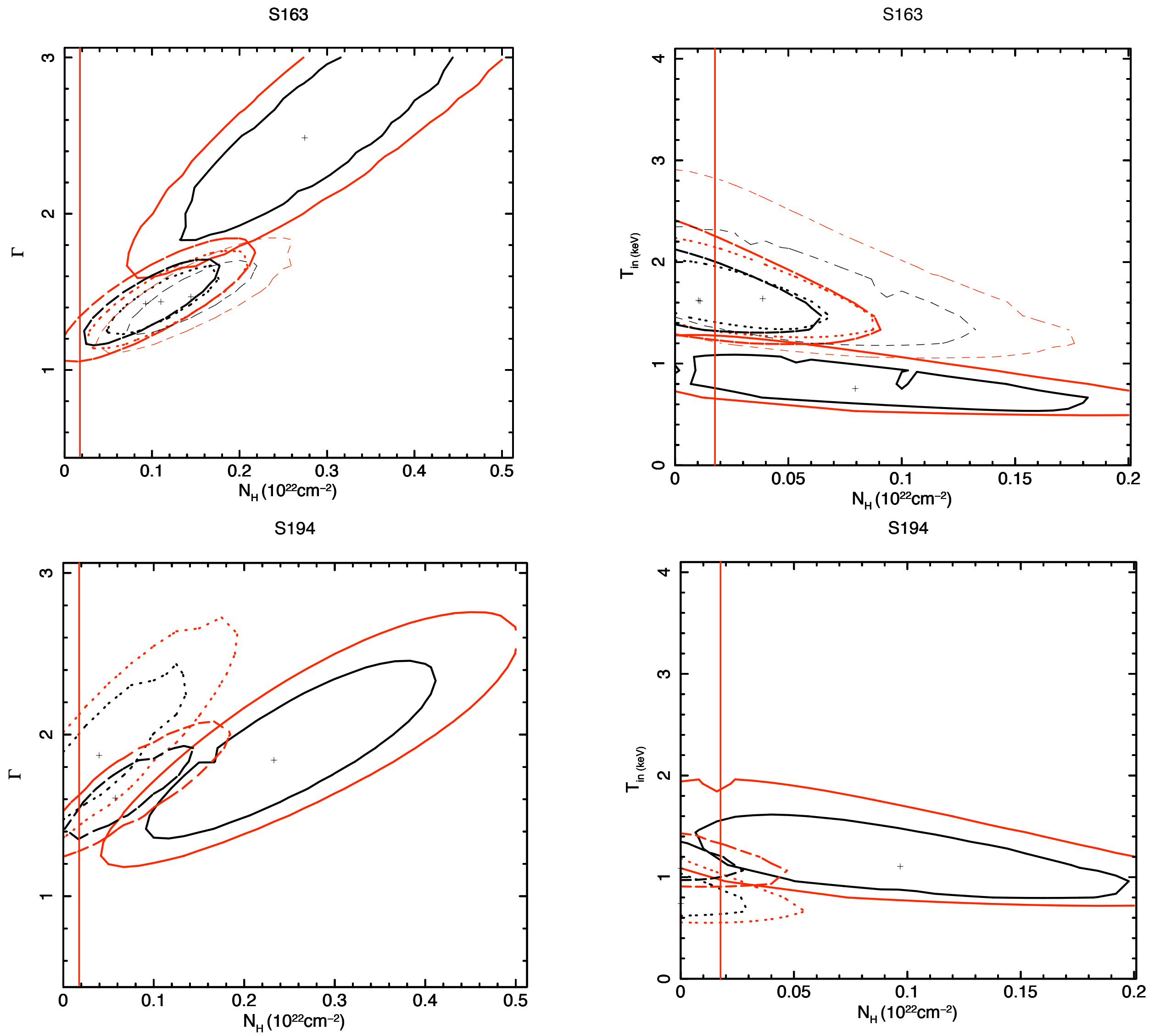

Fig. 6.— $1\sigma$ and $2\sigma$ confidence contours for two interesting fit parameters for S163 (top) and S194 (bottom): left) PO model; right) DBB model. S163: Solid, dotted, dashed, and thin-dashed lines represent the contours from Groups 1, 2, 3 and 4 spectra respectively. S194: Solid, dotted and dashed lines represent the contours from Groups 1, 2 and 3 spectra respectively (see text). The vertical line is the locus of $N_H(Gal.)$. Note that in some instances, e.g. solid contours in S163 power-law fit, $N_H > N_H(Gal.)$ is required, while the same spectrum is consistent with $N_H(Gal.)$ in the DBB fit; in the assumption of composite PO + DBB source spectra, the B10 simulations suggest that this spectrum contains or is dominated by a DBB component - see fig. 8.

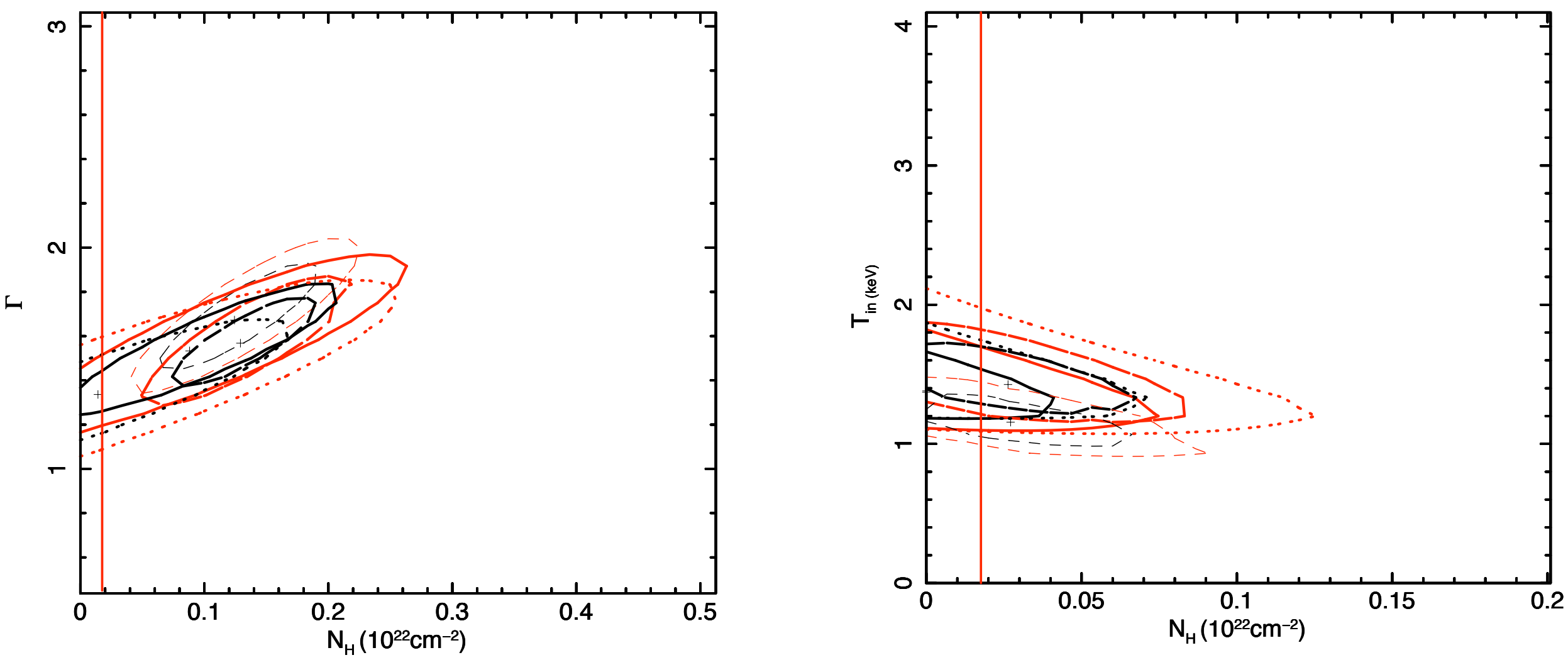

Fig. 7.— 1$\sigma$ and 2$\sigma$ confidence contours for two interesting fit parameters for coadded spectra of individual GC sources: left) PO model ; right) DBB model. Solid, dotted, dashed and thin dashed lines represent the contours for S96, S194, S163 and S185 respectively. The vertical line is the locus of $N_H(Gal.)$.

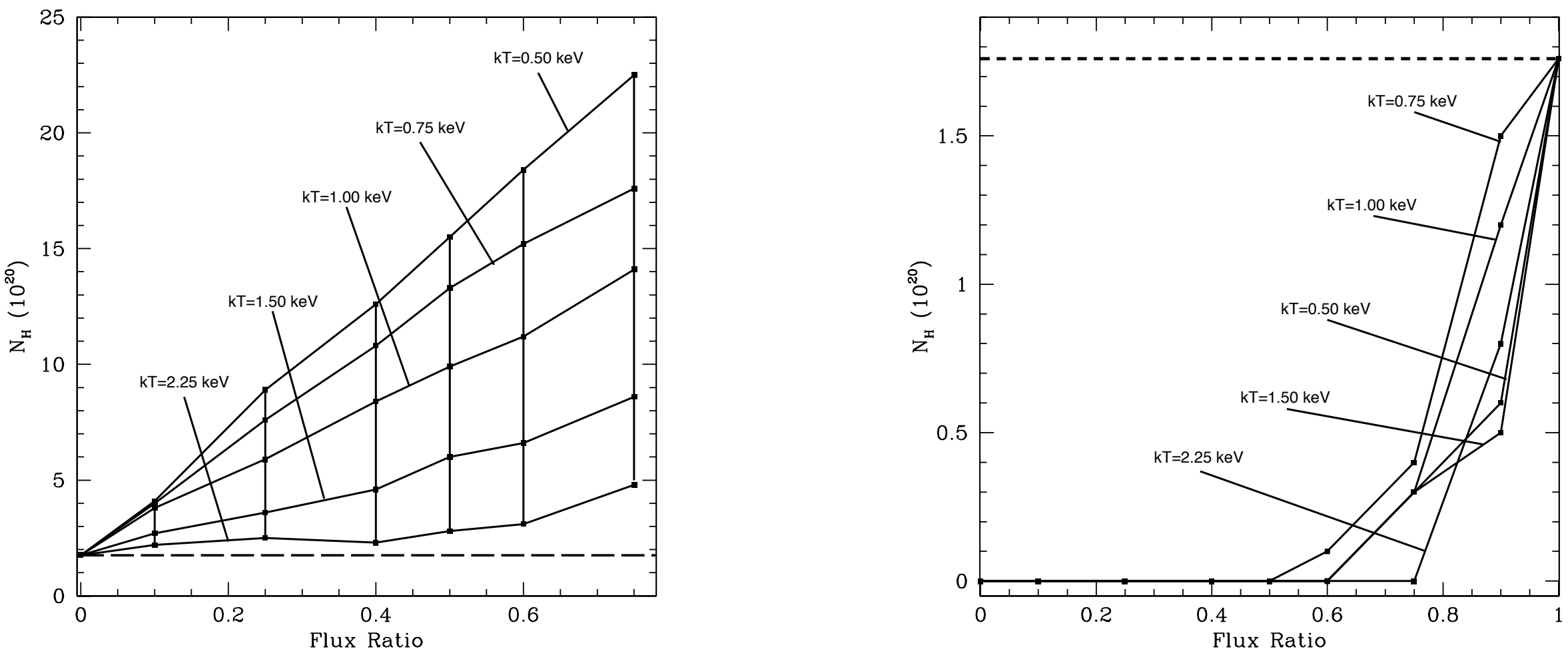

Fig. 8.— Simulations of the results of fitting composite power-law + disk emission spectra with single emission models, from B10. Left: Best-fit $N_H$ for a power-law only fit versus the ratio of the disk/total fluxes for a range of disk temperatures; the dashed horizonthal line represents the Galactic $N_H$ along the line of sight to NGC 4278. Right: the same but for a disk only fit. A power-law $\Gamma = 1.7$ was used for the simulated spectra, however compatible results were found for a range of $\Gamma = 1.5 - 2.5$, see B10.

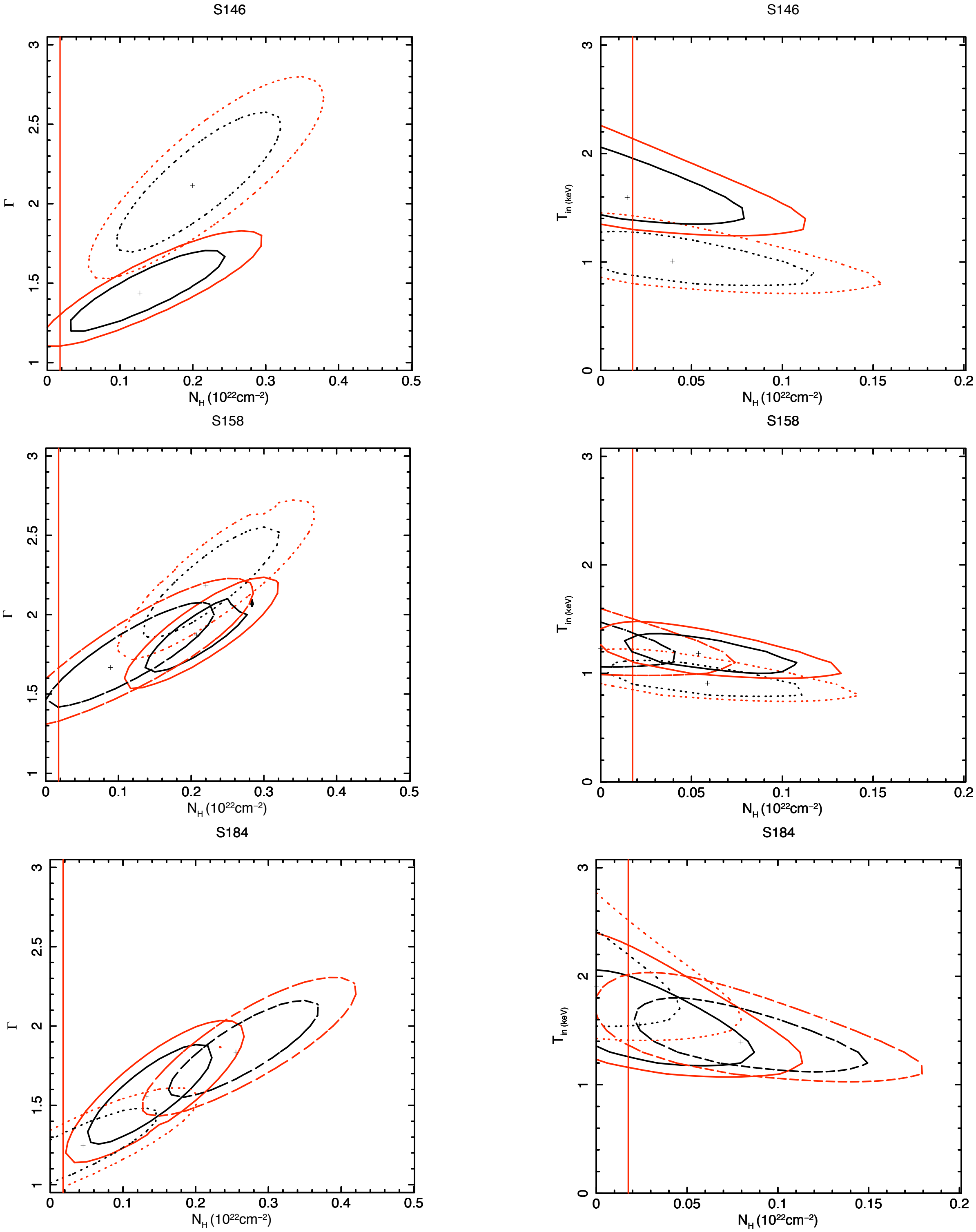

Fig. 9.— 1$\sigma$ and 2$\sigma$ confidence contours for two interesting fit parameters for the field sources S146 (top), S158 (middle) and S184 (bottom): left) PO model; right) DBB model. Solid, dotted and dashed lines represent the contours from Groups 1, 2 (and 3 when available,) spectra. The vertical line is the locus of $N_H(Gal.)$. As suggested by the B10 simulations - see fig.8 - a spectrum with a DBB component will result in $N_H > N_H(Gal.)$ when fitted with a single power-law model.

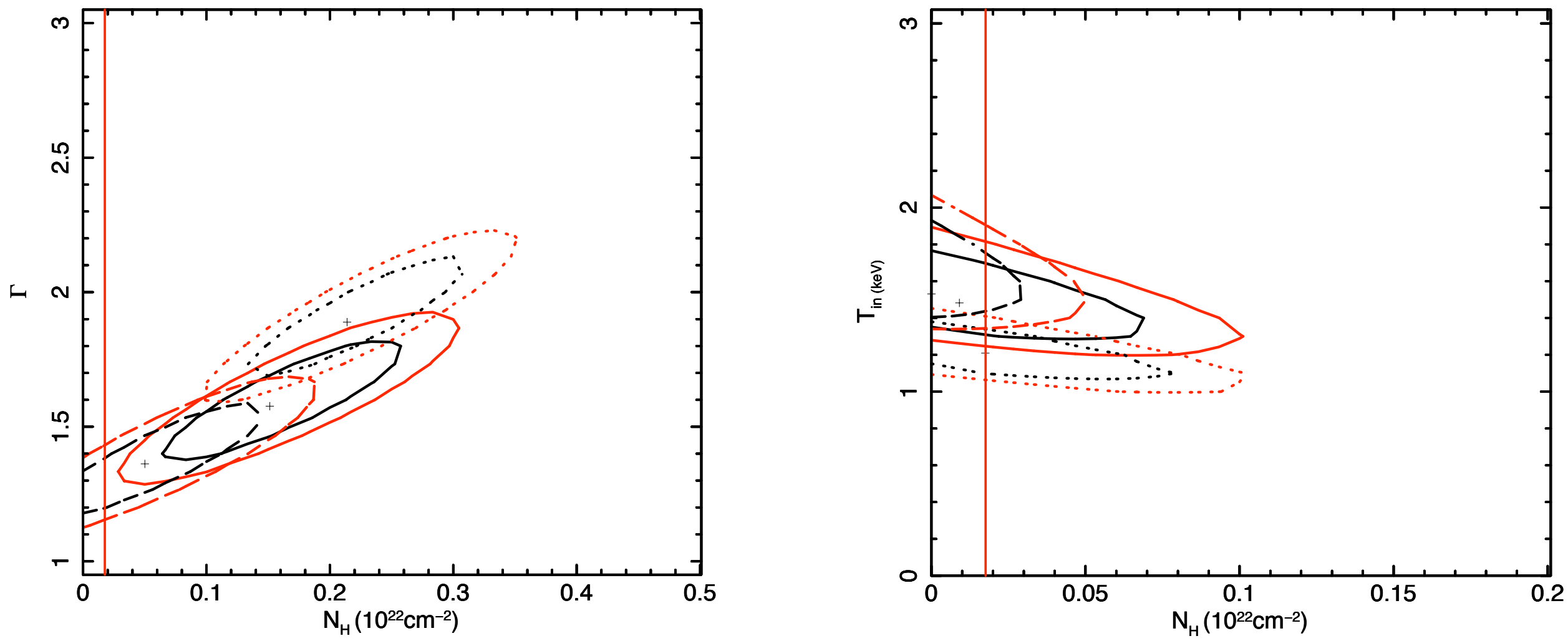

Fig. 10.— $1\sigma$ and $2\sigma$ confidence contours for two interesting fit parameters for the coadded individual field sources: left) PO model; right) DBB model. Solid, dotted and dashed lines represent the contours for S146, S158 and S184 respectively. The vertical line is the locus of $N_H(Gal.)$.

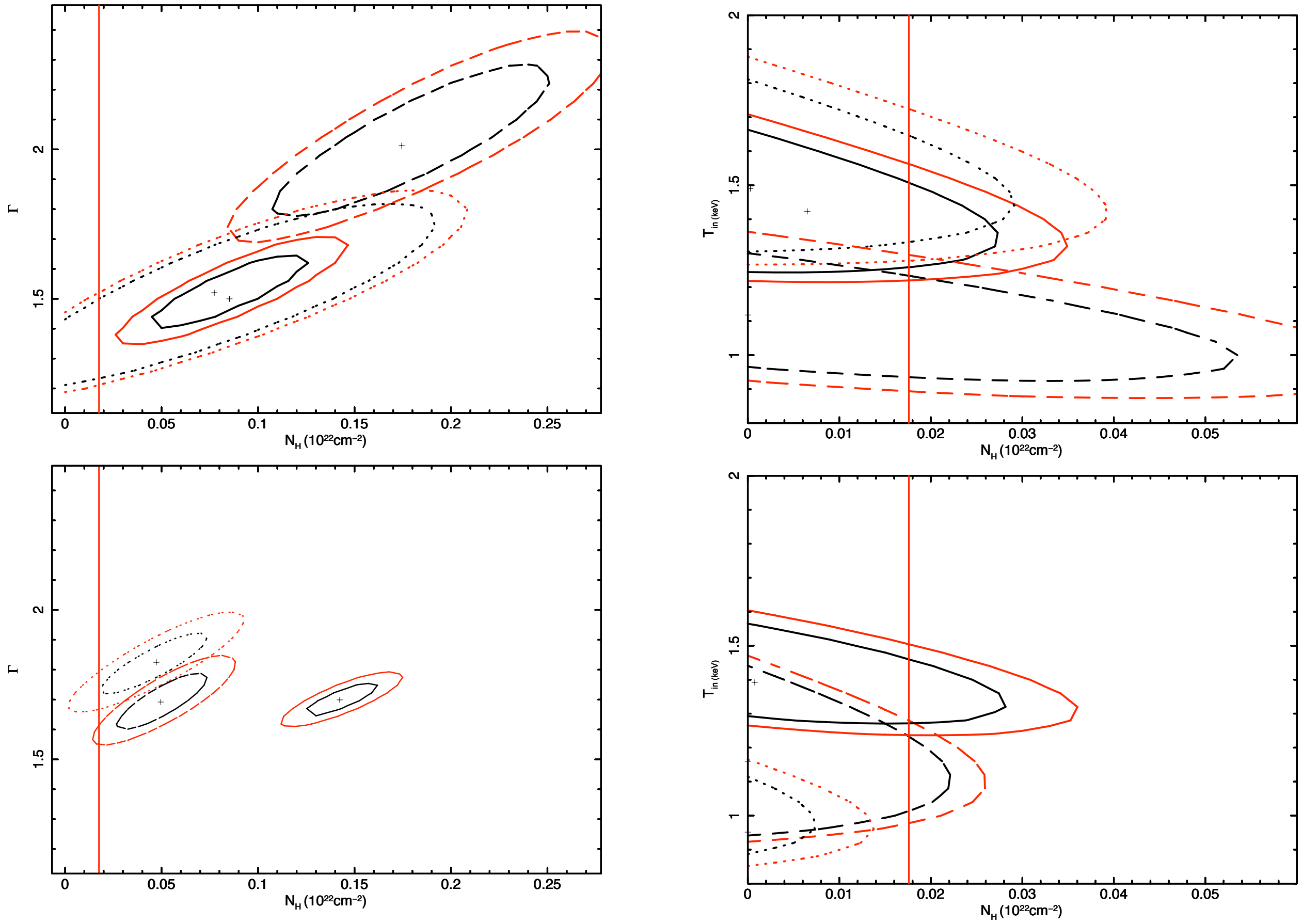

Fig. 11.— 1$\sigma$ and 2$\sigma$ confidence contours for two interesting fit parameters for GC (top) and Field (bottom) luminosity sample sources: Left) PO model; Right) DBB model. Solid, dotted and dashed lines represent the contours from the high, mid and low-luminosity sample spectra respectively. The vertical lines represent the locus of $N_H(Gal.)$. Based on the B10 simulations, the high-luminosity contours are consistent with DBB dominated emission for both GC and field LMXBs. We find, however, a possible difference in the spectral results of the low-luminosity field and GC samples, suggesting higher $N_H$, in excess of the Galactic line of sight column, in the latter, while the former could be fitted with $N_H(Gal.)$.

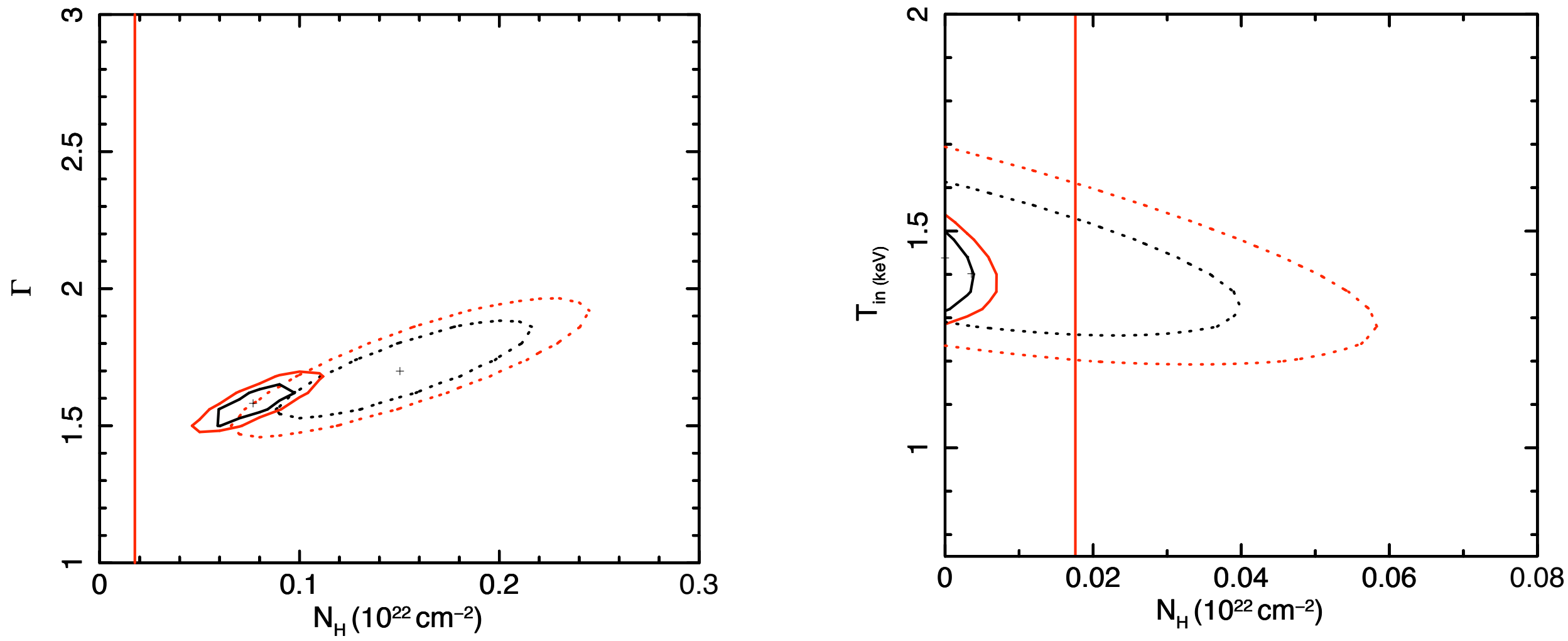

Fig. 12.— 1$\sigma$ and 2$\sigma$ confidence contours for GC color samples: Left) PO model; Right) DBB model. Solid and dotted lines represent the contours for the red and blue sample respectively. The vertical lines represent the locus of $N_H(Gal.)$.

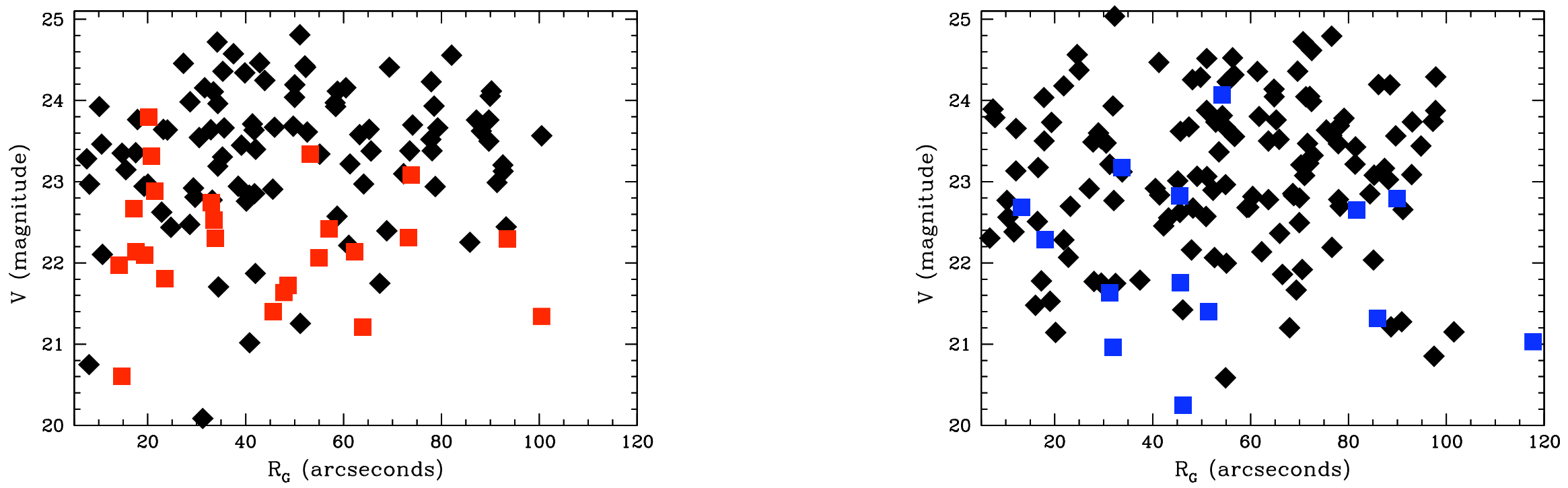

Fig. 13.— $V$ band magnitude against galactocentric distance $R_G$ for all red (left) and blue (right) GCs. The colored red and blue squares in each plot indicate that an X-ray source was present in the GC. The diamonds represent GCs with no X-ray counterpart.There is a trend for GCs to host LMXBs with V, but not $R_G$.

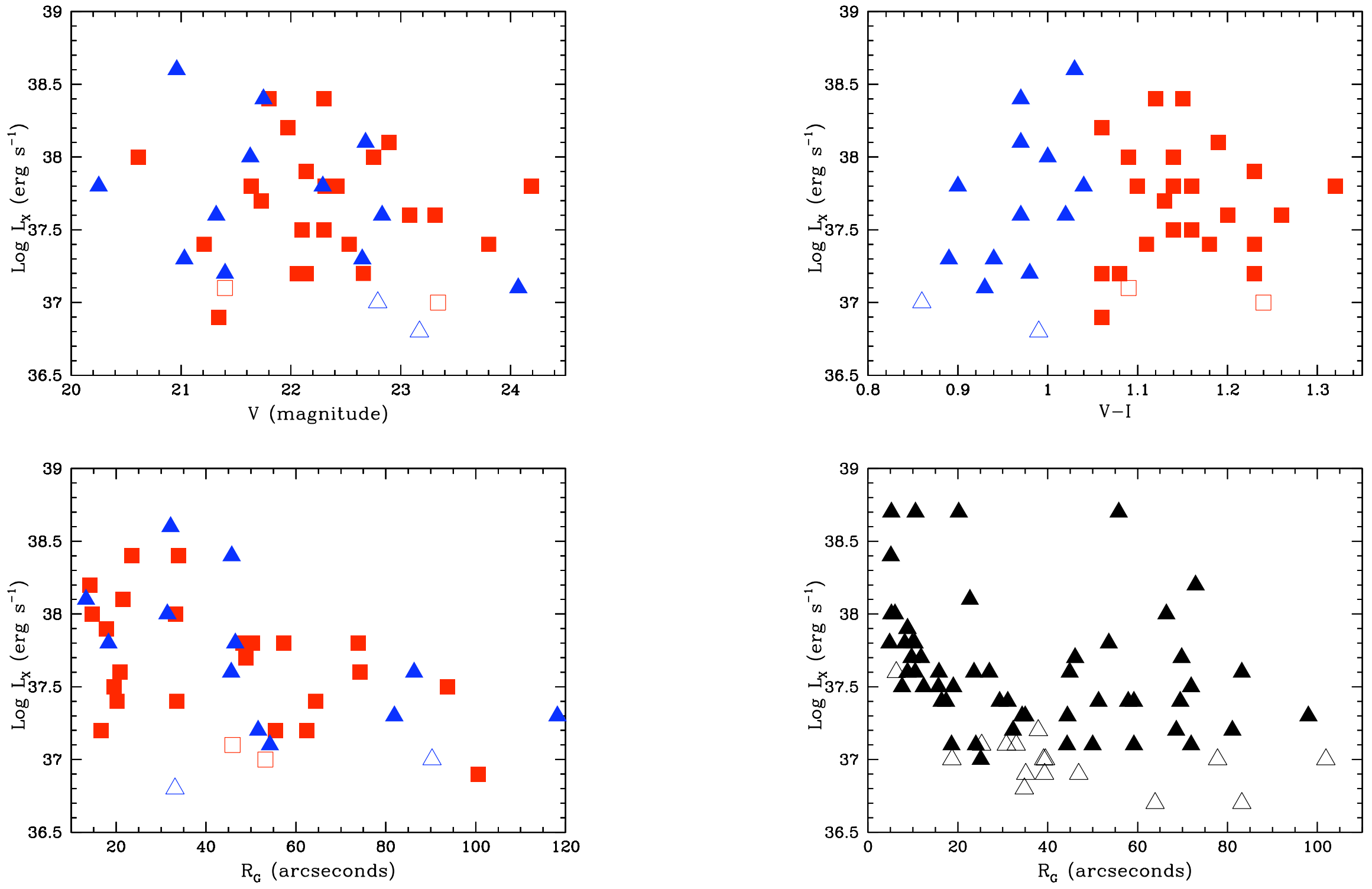

Fig. 14.— Top: Left) scatter plots of GC $L_X$ versus $V$; Right) $L_X$ versus $V-I$. Bottom: Left) $L_X$ versus $R_G$ for GC LMXBs; Right) $L_X$ versus $R_G$ for field LMXBs (this includes points within 10” which were not used in the analysis - see text). In the GC scatter plots, red and blue GC are indicated by red and blue points, respectively. Filled symbols identify the $> 3\sigma$ samples. As discussed in the text, the $L_X - R_G$ correlation in the GC-LMXB sample is highly significant.

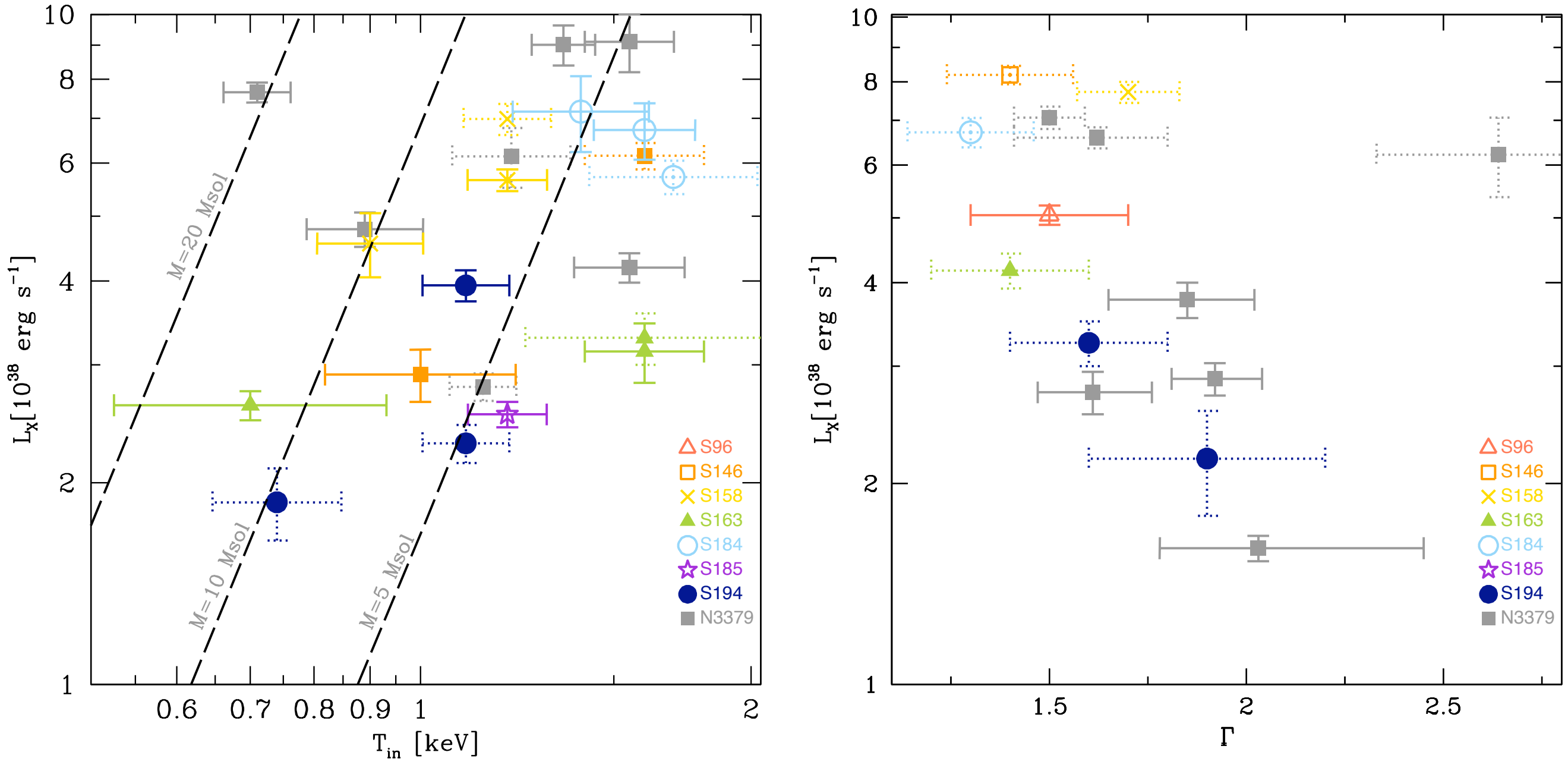

Fig. 15.— Left: $L_X - kT_{in}$, Right: $L_X - \Gamma$. diagrams for the NGC 4278 luminous LMXBs. Each point represents the results of a spectral fit and $1\sigma$ error, as described in the text. The results for a given source are displayed with the same symbols, as listed in the $L_X - \Gamma$ figure. Cases in which either thermal or power-law results are acceptable are displayed in both diagrams with dot-dashed lines. The analogous results of B10 for NGC3379 are plotted in gray. Dotted lines represent cases of possible intermediate spectra, see B10. The diagonal lines in the $L_X - kT_{in}$ plot represent loci of BH mass from eqn. 3 of Gierlinski & Done 2004 (see B10 for details). The ranges of $\Gamma$, $kT_{in}$, and BH masses we derive in NGC 4278 are all consistent with those of Galactic BHBs (see text).

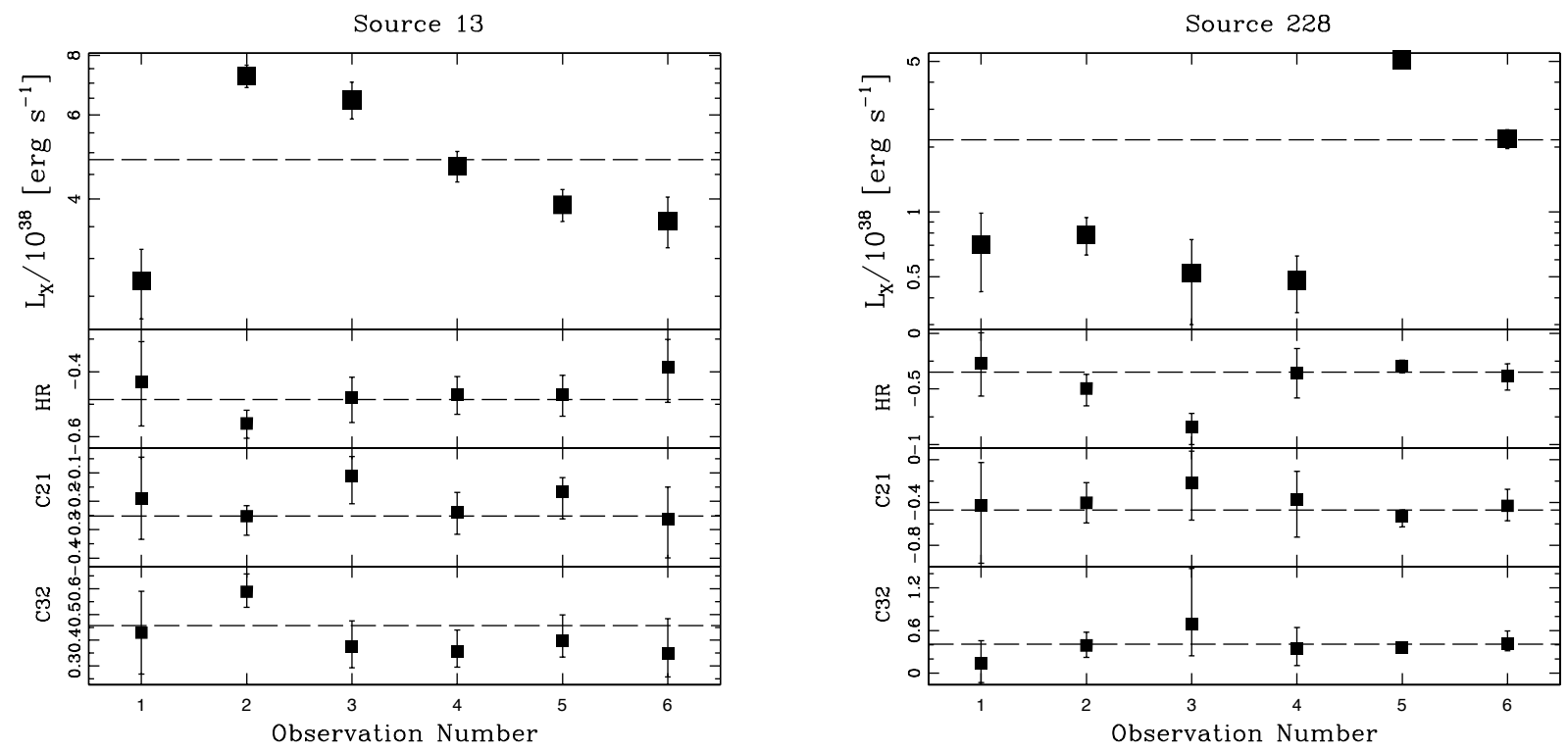

Fig. 16.— Light curves of S13 and S228 from B09, indicating luminosity and spectral variability. Each point represents a separate observation. From the top down: (0.3-8.0) keV $L_X$; hardness ratio, defined as $HR = H - S/(H+S)$ where $H$ is the number of counts in the hard band (2.0-8.0) keV and S is the number of counts in the soft band (0.5-2.0) keV; the color $C21 = logS2 + logS1$; and the color $C32 = logS2 - logH$, where $S21 = 0.3 - 0.9$ keV, $S2 = 0.9 - 2.5$ keV and $H = 2.5 - 8.0 keV$ (see B09)

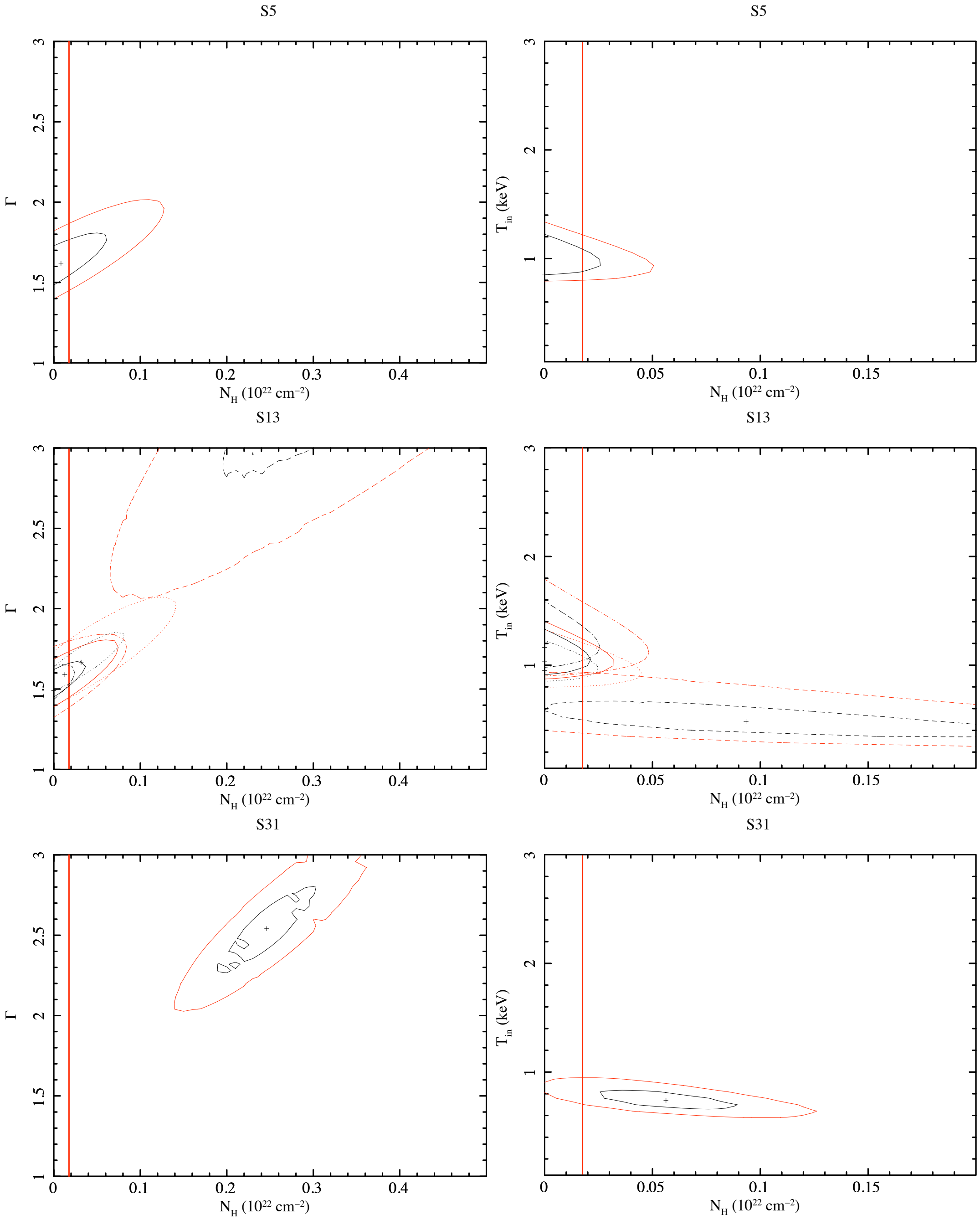

Fig. 17.— Luminous sources outside the *HST* field of view: 1$\sigma$ (black) and 2$\sigma$ (red) confidence contours from the power law (left) and DiskBB (right) models, using All Obs (solid contours), Group 1 Obs (dashed contours), Group 2 Obs (dotted contours) and Group 3 Obs (dot-dashed contours) (where applicable).

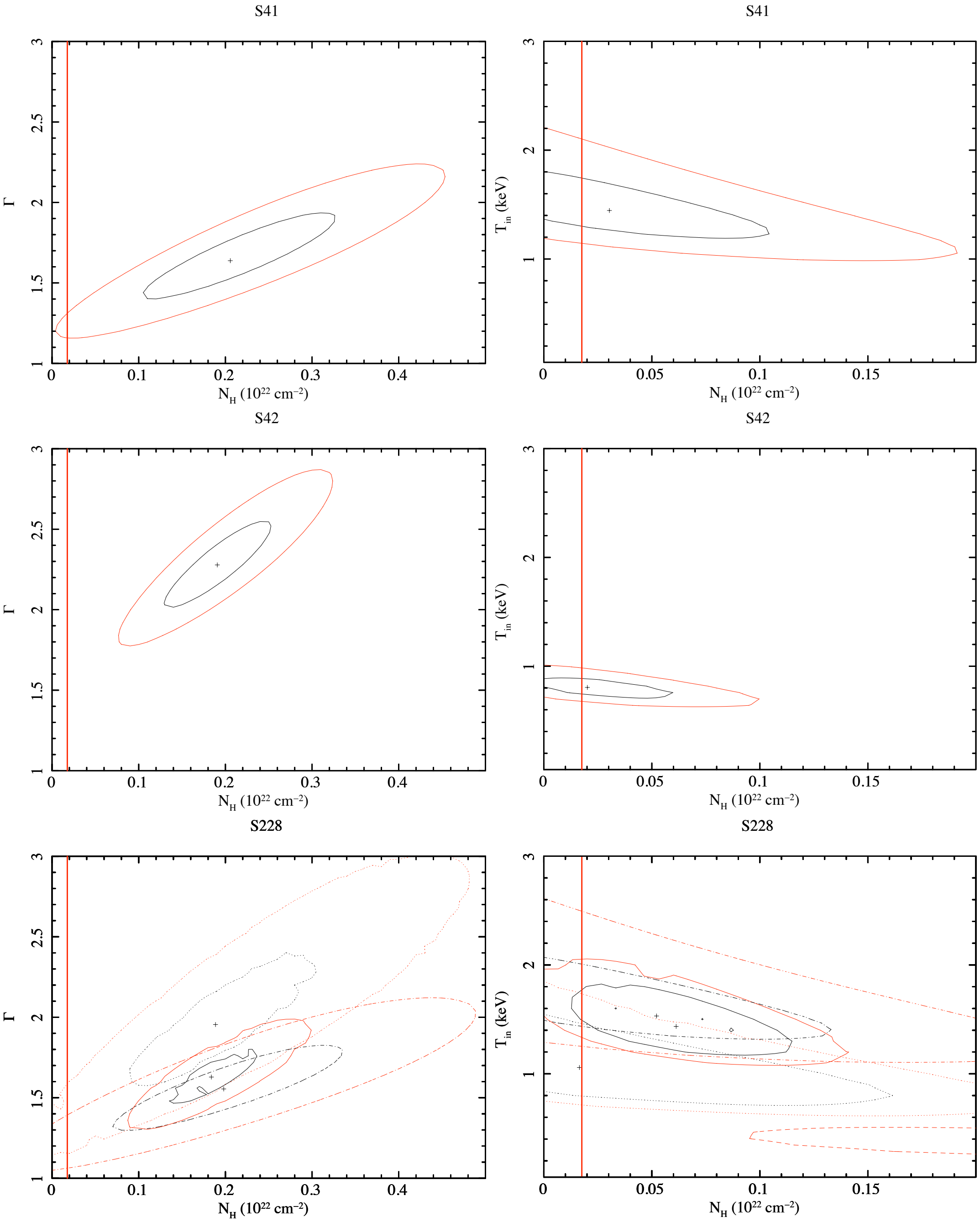

Fig. 18.— Luminous sources outside the *HST* field of view: 1$\sigma$ (black) and 2$\sigma$ (red) confidence contours from the power law (left) and DiskBB (right) models, using All Obs (solid contours), Group 1 Obs (dashed contours), Group 2 Obs (dotted contours) and Group 3 Obs (dot-dashed contours) (where applicable).